\documentclass[twocolumn]{aastex631}

\begin{document}
%\linenumbers
\title{Ultraviolet flux and spectral variability study of blazars observed with UVIT/\textit{AstroSat}}
\correspondingauthor{C. S. Stalin; S. B. Gudennavar}
\email{stalin@iiap.res.in; shivappa.b.gudennavar@christuniversity.in}
\author[0000-0002-0786-7307]{M. Reshma}
\affiliation{Department of Physics and Electronics, CHRIST University, Bangalore 560 029, India}

\author[0000-0002-0786-7307]{Aditi Agarwal}
\affiliation{Centre for Cosmology and Science Popularization, SGT University, Gurugram 122 505, India}

\author[0000-0002-0786-7307]{C. S. Stalin}
\affiliation{Indian Institute of Astrophysics, Block II, Koramangala, Bangalore 560 034, India}

\author[0000-0002-0786-7307]{Prajwel Joseph}
\affiliation{Indian Institute of Astrophysics, Block II, Koramangala, Bangalore 560 034, India}
\affiliation{Department of Physics and Electronics, CHRIST University, Bangalore 560 029, India}

\author[0000-0002-0786-7307]{Akanksha Dagore}
\affiliation{Indian Institute of Astrophysics, Block II, Koramangala, Bangalore 560 034, India}

\author[0000-0002-0786-7307]{Amit Kumar Mandal}
\affiliation{Department of Physics $\&$ Astronomy, Seoul National University, Seoul 08826, Republic of Korea}

\author[0000-0002-0786-7307]{Ashish Devaraj}
\affiliation{Department of Physics and Electronics, CHRIST University, Bangalore 560 029, India}

\author[0000-0002-0786-7307]{S. B. Gudennavar}
\affiliation{Department of Physics and Electronics, CHRIST University, Bangalore 560 029, India}
\begin{abstract}
Blazars, the peculiar class of active galactic nuclei (AGN), are known to show flux variations across the accessible electromagnetic spectrum. Though they have been studied extensively for their flux variability characteristics across wavelengths, information on their ultraviolet (UV) flux variations on time scales of hours is very limited. Here, we present the first UV flux variability study on intraday time scales of a sample of 10 blazars comprising 2 flat spectrum radio quasars (FSRQs) and 8 BL Lacertae objects (BL Lacs). These objects, spanning a redshift ($z$) range of 0.034 $\le$ $z$ $\le$ 1.003, were observed in the far-UV (FUV: 1300 $-$ 1800 \AA) and near-UV (NUV: 2000 $-$ 3000 \AA) wavebands using the ultraviolet imaging telescope on board \textit{AstroSat}. UV flux variations on time scales of hours were detected in 9 sources out of the observed 10 blazars. The spectral variability analysis showed a bluer-when-brighter trend with no difference in the UV spectral variability behavior between the studied sample of FSRQs and BL Lacs. The observed UV flux and spectral variability in our sample of both FSRQs and BL Lacs revealed that the observed UV emission in them is dominated by jet synchrotron process.
\end{abstract}

\keywords{Active galactic nuclei (16) - Blazars (164) - BL Lac objects (168) - flat spectrum radio quasars (2163) - Ultraviolet photometry (1740)}

\section{Introduction} \label{sec:intro}
Active Galactic Nuclei (AGN), the high luminosity (10$^{11}$ $-$ 10$^{14}$ L$_{\odot}$) sources in the Universe, are thought to be powered by the accretion of matter onto super massive black holes (SMBHs: 10$^6$ $-$ 10$^{10}$ M$_{\odot}$) situated at the center of galaxies \citep{1969Natur.223..690L,1984ARA&A..22..471R}. About 10\% of AGN emit copiously in the radio band, display relativistic jets, and emit over a wide range of the electromagnetic spectrum from low energy radio waves to high energy (TeV) $\gamma$-rays. A minority of these objects having their relativistic jets oriented close to the line of sight to the observer ($\le$ 10$^{\circ}$)  are called blazars \citep{1993ARA&A..31..473A,1995PASP..107..803U}. The broad band spectral energy distribution (SED) of blazars is dominated by the beamed emission from relativistic particles in their jet \citep{1978PhyS...17..265B,1995PASP..107..803U}. 
\\[6pt]
Blazars are divided into flat spectrum radio quasars (FSRQs) and BL Lacertae objects (BL Lacs), with FSRQs having broad emission lines with equivalent width above 5 \AA\ and BL Lacs having either a featureless spectra or spectra with weak emission lines with equivalent width below 5 \AA. A more physical distinction between FSRQs and BL Lacs is based on the luminosity of the broad line region ($L_{BLR})$ relative to the Eddington luminosity ($L_{Edd}$) with the dividing line set at $L_{BLR}/L_{Edd}$ $\sim$ 5 $\times$ 10$^{-4}$ \citep{2011MNRAS.414.2674G}. The broad band SED of blazars has a two hump structure. The low energy hump peaking in the optical/infrared/soft X-ray region is attributed to synchrotron emission process \citep{1982ApJ...253...38U} and the high energy hump  peaking in the X-ray/MeV region is attributed to inverse  Compton scattering process \citep{2010ApJ...716...30A}. Based on the position of the synchrotron peak ($\nu_{peak}$) in their broad band SED, blazars are further divided into low synchrotron peaked (LSP; $\nu_{peak} < 10^{14}$ Hz), intermediate synchrotron peaked (ISP; $10^{14}$ Hz $< \nu_{peak} < 10^{15}$ Hz) and high synchrotron peaked (HSP; $\nu_{peak} > 10^{15}$ Hz) blazars \citep{2010ApJ...716...30A}. 
\\[6pt]
In addition, an important characteristic of blazars is that they show flux variations over the entire accessible electromagnetic spectrum on a range of time scales from minutes to hours \citep{1995ARA&A..33..163W,1997ARA&A..35..445U,webb202130}. Such flux variations can serve as an efficient tool to understand the nature of the central regions and the jets of blazars. The flux variations are also known to be correlated across wavelengths, supporting the argument that the low energy and the high energy emission arises from the same population of relativistic electrons in the jet via synchrotron and inverse Compton processes, respectively \citep{1997ARA&A..35..445U}. However, recent observations point to varied correlations between low energy optical and high energy $\gamma$-ray observations indicating lack of our understanding on the flux variability characteristics of blazars \citep{2020MNRAS.498.5128R,2021MNRAS.504.1772R}. In addition to flux variations, blazars also show large optical and infrared polarization \citep{1980ARA&A..18..321A} and optical polarization variations \citep{2017ApJ...835..275R,2022MNRAS.510.1809P,2022MNRAS.517.3236R}. Though blazars have been studied for variability across multiple wavelengths, on a range of time scales,  their ultraviolet (UV) variability characteristics are not explored much, with only sparse reports available in the literature \citep{1992ApJ...401..516E}. Understanding the UV flux variations in blazars is important as the UV emission in the broad band SED of blazars is usually dominated by synchrotron emission from relativistic jet electrons \citep[]{abdo2011fermi, paliya2015multi}. However, in the faint state, signature of prominent accretion disk emission in the optical-UV region is evident in the broad band SED of the FSRQ category of blazars \citep[]{bonnoli2011gamma, paliya2016broadband, paliya2017general}. Some recent studies do exist that focus on UV variability of AGN sources, mainly from \textit{Galaxy Evolution Explorer (GALEX)} and \textit{International Ultraviolet Explorer (IUE)} observations \citep{2011A&A...527A..15W,2018JApA...39...65S}. However, these studies are focused on flux variations in non-blazar type AGN on longer time scales.
\\[6pt]
In this work, UV variability characteristics of blazars on hour-like time scales are studied using the data from the ultraviolet imaging telescope (UVIT: \citealt{2017CSci..113..583T,2020AJ....159..158T}) on board \textit{AstroSat}. This is the first study where the temporal and spectral properties of 10 blazars in UV on intraday time scales are investigated. The paper is structured as follows: In Section \ref{sec:Observations and Data Reduction}, observations and data reduction procedures are described. In Section \ref{sec:analysis}, the analysis technique used to study the temporal and spectral variability properties of the sample of sources is detailed, while the notes on individual sources and the results are detailed in Section \ref{sec:results}. Finally, discussion and conclusions are presented in Section \ref{sec:Discussion}.

\begin{table}
\centering
\caption{Details of the sources studied in this work. Here, RA, Dec., and $z$ are the right ascension, declination, and redshift of the sources, respectively \citep{veron2010}. FS refers to FSRQ and BL refers to BL Lac, LSP and HSP refer to low synchrotron peaked and high synchrotron peaked blazars.}
\begin{tabular}{l@{\hspace{0.2cm}}c@{\hspace{0.2cm}}c@{\hspace{0.2cm}}c@{\hspace{0.2cm}}c@{\hspace{0.2cm}}} 
\hline
\hline
Name & RA        & Dec.        & Type /    & $z$ \\
     & (hh:mm:ss)& (dd:mm:ss)  & Subtype  &     \\ 
\hline
PKS 0208$-$512  & 02:10:46.20  & $-$51:01:01.89 & FS/LSP  & 1.003 \\
1ES 0229+200    & 02:32:48.62  &   +20:17:17.48 & BL/HSP  & 0.140 \\
OJ 287          & 08:54:48.88  &   +20:06:30.64 & BL/LSP  & 0.306 \\
1ES 1101$-$232  & 11:03:37.61  & $-$23:29:31.20 & BL/HSP  & 0.186 \\
1ES 1218+304    & 12:21:21.94  &   +30:10:37.16 & BL/HSP  & 0.182 \\ 
H 1426+428      & 14:28:32.61  &   +42:40:21.05 & BL/HSP  & 0.129 \\
PKS 1510$-$089  & 15:12:50.53  & $-$09:05:59.83 & FS/LSP  & 0.360 \\
Mrk 501         & 16:53:52.22  &   +39:45:36.61 & BL/HSP  & 0.034 \\
PKS 2155$-$304  & 21:58:52.07  & $-$30:13:32.12 & BL/HSP  & 0.116 \\ 
1ES 2344+514    & 23:47:04.84 &   +51:42:17.88 & BL/HSP  & 0.044 \\
\hline
\end{tabular}
\label{Table-1}
\end{table}

\begin{deluxetable*}{lcccccc}
\tablecaption{Log of observations. Here, name of the source, date of observation (dd-mm-yyyy), observational ID (OBSID), filters used in the observations, start and end time of the observations in modified julian date (MJD) and the net exposure time in seconds are given in columns 1, 2, 3, 4, 5, 6 and 7, respectively.\label{Table-2}}
\tabletypesize{\footnotesize}
\tablehead{
Name & \colhead{Date of observation} & \colhead{OBSID} & \colhead{Filter} & \colhead{MJD Start} & \colhead{MJD End} & \colhead{Net exposure time}
}
\startdata
PKS 0208$-$512  & 30-10-2016  & A02-114T01-9000000764 & F148W  & 57690.991247  & 57691.386868 &  1337       \\
                &             &                        & F169M  & 57691.588337  & 57691.731484 &  1710       \\ 
                &             &                        & N219M  & 57690.991175  & 57691.524076 &  2236       \\
                &             &                        & N279N  & 57691.588288  & 57691.863146 &  2119       \\  
                & 25-12-2019  & T03-170T01-9000003388  & F148W  & 58841.831743 & 58841.833213 & 119         \\
                &             &                        & F154W  & 58841.904312 & 58841.905736 & 116         \\ 
                &             &                        & F169M  & 58841.976881 & 58842.311886 & 1223        \\
                &             &                        & F172M  & 58842.508345 & 58842.654596 & 1037        \\
1ES 0229+200    & 01-10-2017  & A04-130T01-9000001572  & F154W  & 58027.768812 & 58027.926301 & 4922        \\
                &             &                        & N245M  & 58027.768763 & 58027.926351 & 4951        \\
                & 09-12-2017  & A04-130T01-9000001762  & F154W  & 58096.737591 & 58096.944207 & 4970        \\
                &             &                        & N245M  & 58096.737542 & 58096.944257 & 5005        \\
                & 22-12-2017  & A04-130T01-9000001792  & F154W  & 58108.834869 & 58109.180995 & 3962        \\
                &             &                        & N245M  & 58108.834820 & 58109.109934 & 5046        \\
                & 08-01-2018  & A04-130T01-9000001822  & F154W  & 58126.562901 & 58126.779650 & 4974        \\
                &             &                        & N245M  & 58126.562852 & 58126.779700 & 5006        \\
OJ 287          & 10-04-2017  & T01-163T01-9000001152  & F169M  & 57852.921707 & 57853.686358 & 8422        \\
                &             &                        & N245M  & 57852.921656 & 57853.415465 & 4501        \\
                &             &                        & N263M  & 57853.417144 & 57853.686437 & 3649        \\
                & 18-04-2018  & A04-199T01-9000002040  & F169M  & 58225.694593 & 58227.596754 & 25365       \\
                &             &                        & F172M  & 58223.458444 & 58225.692828 & 28154       \\
                & 18-05-2020  & T03-206T01-9000003672  & F148W  & 58984.390970 & 58984.732999 & 3636        \\
                &             &                        & F154W  & 58984.734732 & 58985.407435 & 4855        \\
                &             &                        & F172M  & 58986.764797 & 58987.710244 & 6497        \\
                &             &                        & F148Wa & 58985.409168 & 58986.561064 & 3782        \\
1ES 1101$-$232  & 30-12-2016  & G06-086T02-9000000936  & F154W  & 57750.400520 & 57752.501664 & 33497       \\
                &             &                        & N263M  & 57750.400471 & 57752.501714 & 33926       \\
1ES 1218+304    & 21-05-2016  & G05-211T01-9000000464  & F148W  & 57528.302435 & 57529.431526 & 13881       \\
                &             &                        & N245M  & 57528.302745 & 57529.431577 & 13847       \\
H 1426+428      & 05-03-2018  & A04-094T01-9000001942  & F148W  & 58181.865544 & 58182.463086 & 6836        \\
                &             &                        & F154W  & 58181.577434 & 58181.863807 & 7940        \\
                &             &                        & N242W  & 58181.577385 & 58181.781955 & 5110        \\
                &             &                        & N245M  & 58181.783577 & 58182.463086 & 9772        \\  
PKS 1510$-$089  & 30-03-2016  & T01-106T01-9000000404  & F172M  & 57476.398455 & 57478.031161 & 26663       \\
                &             &                        & N219M  & 57476.398407 & 57478.031211 & 26866       \\
                & 16-03-2018  & A04-101T01-9000001984  & F172M  & 58192.878269 & 58193.695960 & 13270       \\
                &             &                        & N219M  & 58192.878220 & 58193.696009 & 13349       \\
                & 15-06-2018  & A04-101T01-9000002170  & F172M  & 58284.086589 & 58284.771371 & 12085       \\
Mrk 501         & 15-08-2016  & G05-218T05-9000000602  & F154W  & 57615.260263 & 57615.946790 & 12154       \\
                &             &                        & N219M  & 57615.255611 & 57615.946840 & 12639       \\
                & 28-03-2020  & A07-145T01-9000003594  & F148W  & 58935.087363 & 58937.726873 & 25484       \\
PKS 2155$-$304  & 21-05-2018  & A04-130T03-9000002108  & F154W  & 58259.329918 & 58259.544023 & 2788        \\
                & 24-06-2018  & A04-130T03-9000002186  & F154W  & 58293.495067 & 58293.579373 & 2842        \\
                & 16-09-2019  & A05-117T01-9000003166  & F154W  & 58742.571362 & 58742.777812 & 4117        \\
1ES 2344+514    & 06-06-2017  & A03-033T01-9000001276  & F172M  & 57910.717527 & 57910.863068 & 2195        \\
                &             &                        & N245M  & 57910.717477 & 57910.860639 & 2015        \\
                & 09-07-2017  & A03-033T01-9000001368  & F172M  & 57943.391924 & 57943.546931 & 3586        \\
                &             &                        & N245M  & 57943.391873 & 57943.549083 & 3718        \\ 
                & 07-08-2017  & A03-033T01-9000001438  & F172M  & 57972.282651 & 57972.430529 & 3886        \\
                &             &                        & N245M  & 57972.282601 & 57972.428121 & 3705        \\
                & 07-09-2017  & A03-033T01-9000001524  & F172M  & 58003.815110 & 58003.972611 & 3977        \\
                &             &                        & N245M  & 58003.815060 & 58003.972663 & 4000        \\
                & 22-11-2017  & A04-049T01-9000001710  & N245M  & 58078.933083 & 58079.201251 & 1897        \\
                & 07-12-2017  & A04-049T01-9000001754  & F172M  & 58094.972718 & 58095.243561 & 1392        \\
                &             &                        & N245M  & 58094.972670 & 58095.243611 & 1409        \\
                & 12-09-2018  & T02-104T01-9000002356  & F172M  & 58372.864037 & 58373.064969 & 3439        \\                
\enddata
\end{deluxetable*}

\begin{table}
\centering
\caption{Details of FUV and NUV filters of UVIT used in this work. Here, $\lambda_{mean}$ is the mean wavelength and $\Delta \lambda$ is the bandwidth \citep{2020AJ....159..158T}.}
\begin{tabular}{l@{\hspace{1cm}}c@{\hspace{1cm}}r@{\hspace{1cm}}c@{\hspace{1cm}}} 
\hline
\hline
Filter & $\lambda_{mean}$ & $\Delta \lambda$ & Zero point \\
       & (\AA)   &  (\AA)   & (mag)  \\
\hline
F148W   &  1481            &  500              & 18.097 $\pm$ 0.010      \\
F148Wa  &  1485            &  500              & 18.097 $\pm$ 0.010      \\
F154W   &  1541            &  380              & 17.771 $\pm$ 0.010      \\
F169M   &  1608            &  290              & 17.410 $\pm$ 0.010      \\
F172M   &  1717            &  125              & 16.274 $\pm$ 0.020      \\
N242W   &  2418            &  785              & 19.763 $\pm$ 0.002     \\
N219M   &  2196            &  270              & 16.654 $\pm$ 0.020      \\
N245M   &  2447            &  280              & 18.452 $\pm$ 0.005      \\
N263M   &  2632            &  275              & 18.146 $\pm$ 0.010      \\
N279N   &  2792            &  90               & 16.416 $\pm$ 0.010      \\
\hline
\end{tabular}
\label{Table-3}
\end{table}

\section{Observations and Data Reduction}
\label{sec:Observations and Data Reduction}
\begin{figure*}
\hbox{
\includegraphics[scale=0.30]{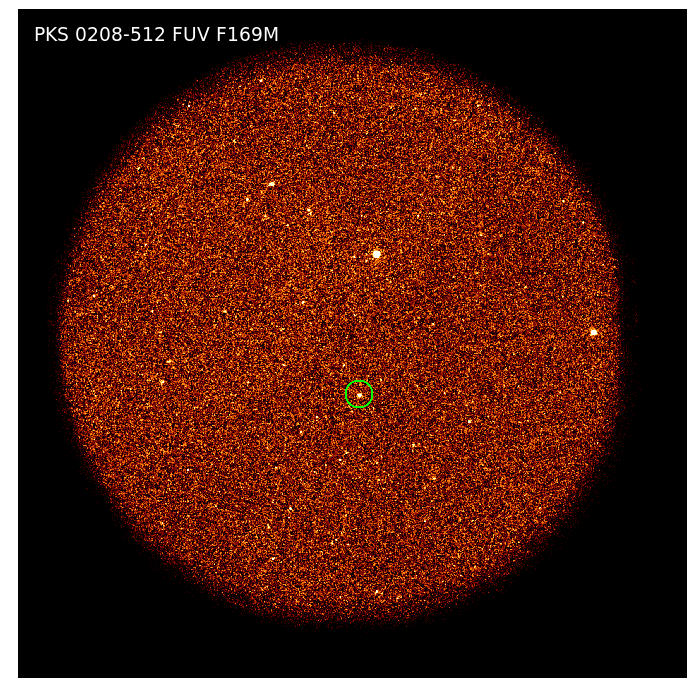}
\includegraphics[width=5.7cm,height=5.6cm]{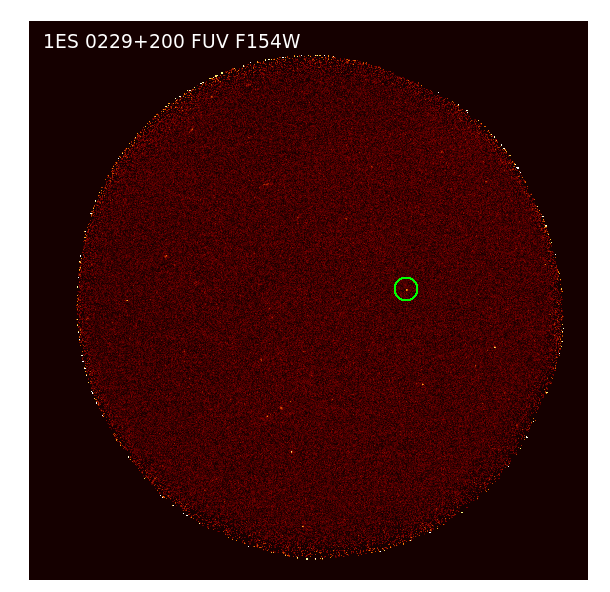}
\includegraphics[scale=0.30]{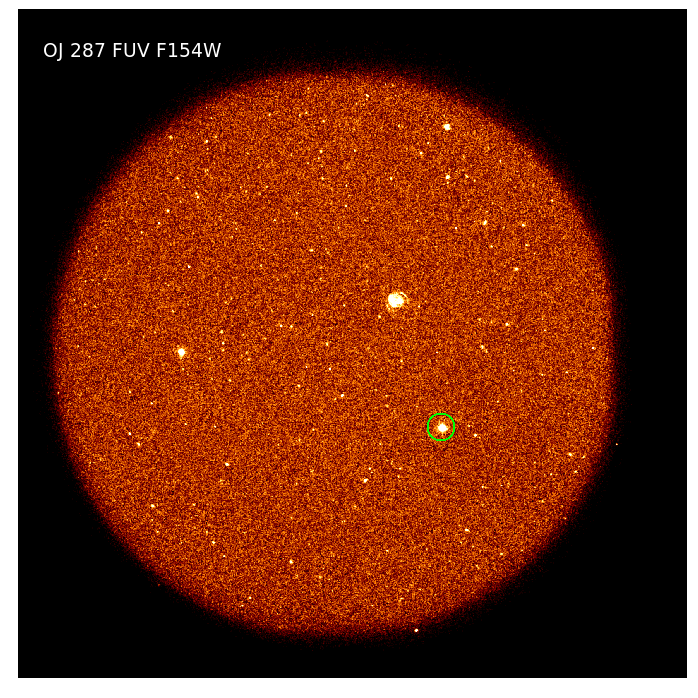}
}

\hbox{
\includegraphics[scale=0.30]{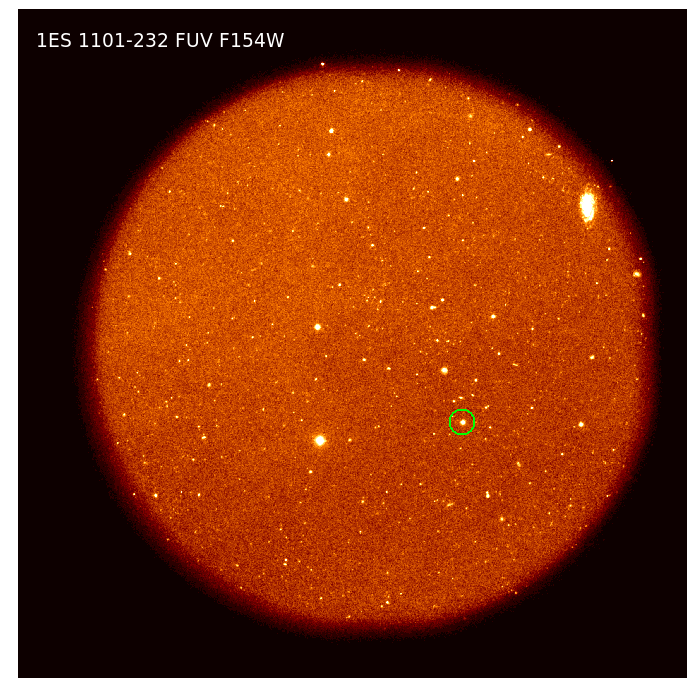}
\includegraphics[scale=0.30]{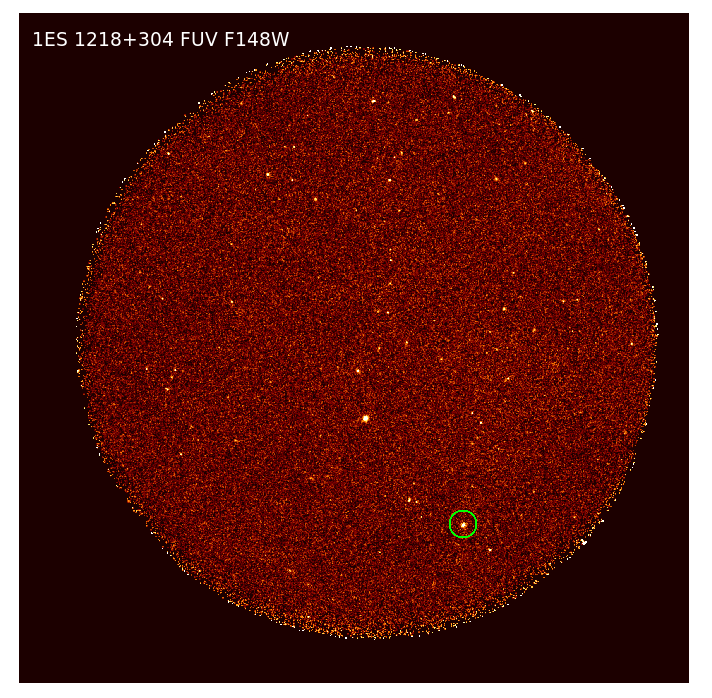}
\includegraphics[scale=0.30]{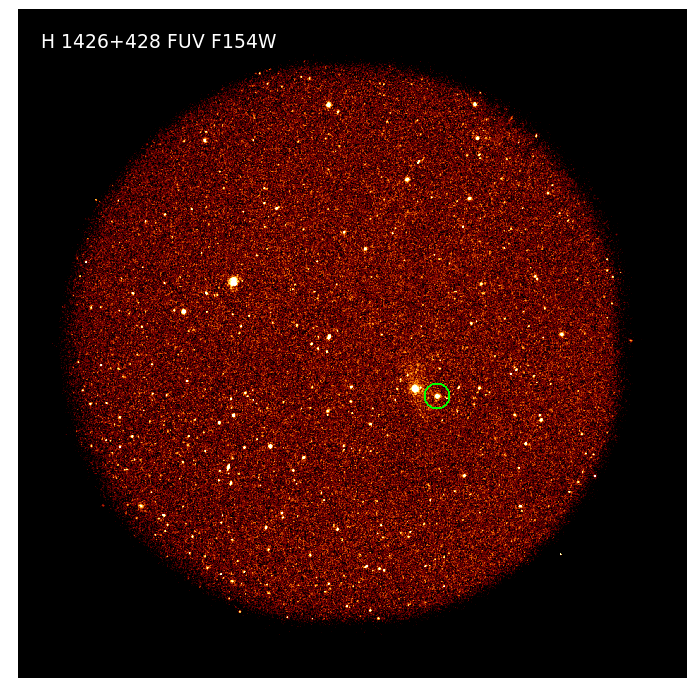}
}

\hbox{
\includegraphics[scale=0.30]{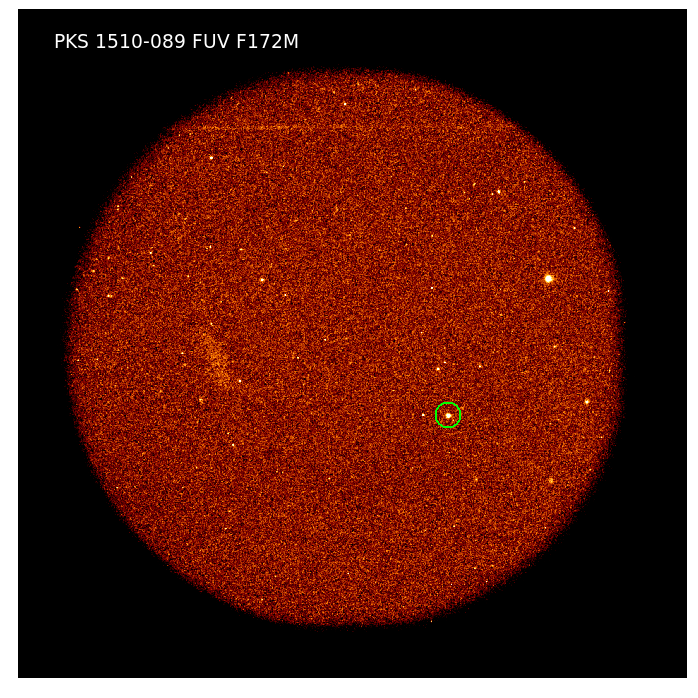}
\includegraphics[scale=0.30]{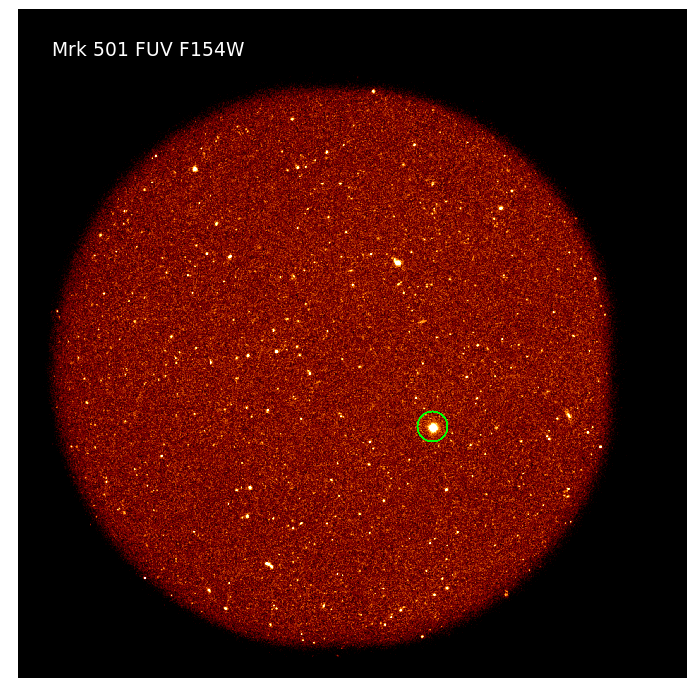}
\includegraphics[scale=0.30]{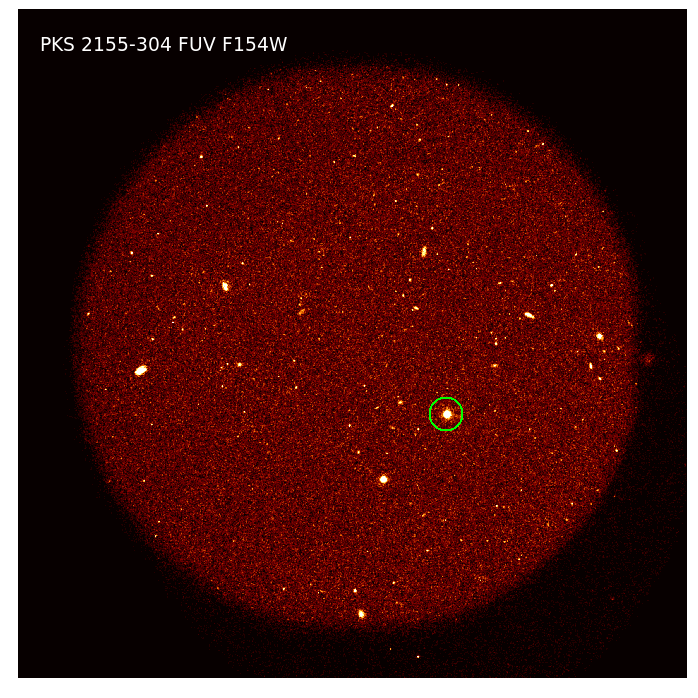}
}

\hbox{
\includegraphics[scale=0.30]{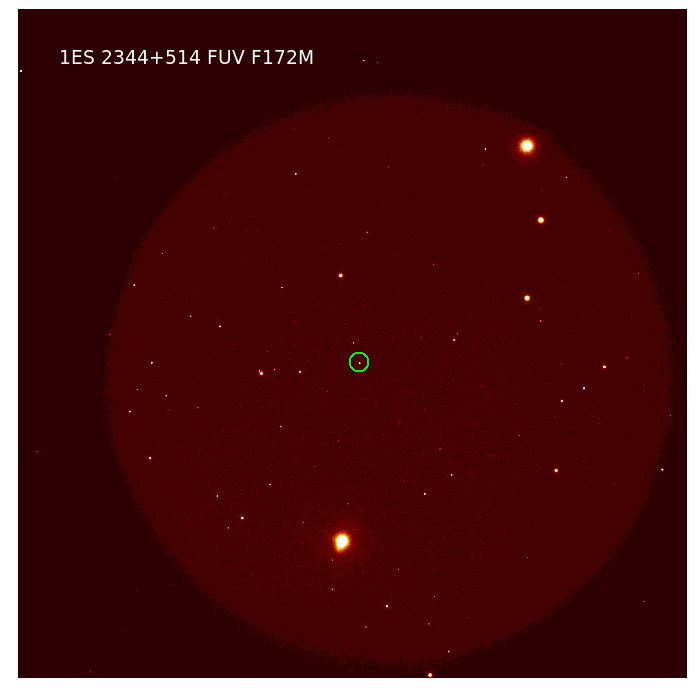}
}

\caption{The observed FUV field images of the ten blazar sources. The target sources are shown in green circles. The source name and the corresponding filter name are given in each field image. Each image has a field of view of $\sim$ 28 arcmin diameter.}
\label{figure-1}
\end{figure*}

\begin{figure}
\centering
\includegraphics[scale=0.80]{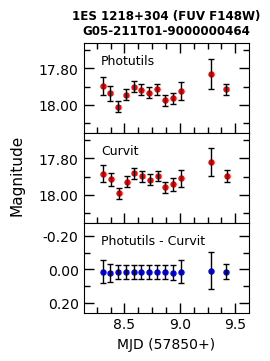}
\caption{Comparison plot of FUV lightcurve for the source 1ES 1218+304 in F148W filter with top panel denoting {\it Photutils} lightcurve, middle panel denoting {\it Curvit} lightcurve, and bottom panel denoting the difference between {\it Photutils} and {\it Curvit} lightcurves. }
\label{figure-2}
\end{figure}

The sample of sources used for this study was selected from the archives of observations carried out by UVIT (\citealt{2017CSci..113..583T,2020AJ....159..158T}), one of the
payloads on board India's multi-wavelength astronomical observatory, \textit{AstroSat} \citep{2017JApA...38...27A}, launched by the Indian Space Research Organization on 28 September 2015. UVIT with a field of view of $\sim$ 28 arcmin diameter observes simultaneously in two channels namely, the far-UV (FUV: 1300 $-$ 1800 \AA) and the near-UV (NUV: 2000 $-$ 3000 \AA) using a set of filters. It also has a visual (VIS) channel (3200 $-$ 5500 \AA), which is used for tracking the aspects of the telescope, and is used in the processing of the data while generating science-ready images in the FUV and NUV channels. The details of the sources studied in this work are given in Table \ref{Table-1}. The log of observations of the sources used in this work is given in Table \ref{Table-2} and the details of the filters used are given in Table \ref{Table-3}. The science-ready images of the observations available at the Indian Space Science Data Center (ISSDC) \footnote{https://www.issdc.gov.in/astro.html} are used in this work. The images were made available to ISSDC by the Payload Operations Center (POC) at the Indian Institute of Astrophysics, Bangalore. At the POC, the images were reduced using the UVIT L2 pipeline version 6.3 \citep{2021JApA...42...29G,2022JApA...43...77G}. The pipeline creates science-ready images by correcting the raw data for spacecraft drift, flat field, and geometric distortion. The observed fields of the sources are given in Fig. \ref{figure-1}. The aperture photometry for one of the target sources, 1ES 1218+304, for a radius of 12 sub-pixels was carried out using the {\it Photutils} Python package \citep{larry_bradley_2023_1035865}. Background was removed using the local background subtraction method, where a circular annular region with an inner radius of 150 sub-pixels and an outer radius of 200 sub-pixels were used. The background-subtracted source light curve obtained using {\it Photutils} was compared with the light curve generated using the {\it Curvit} Python package \citep{2021JApA...42...25J} (Fig. \ref{figure-2}). From Fig. \ref{figure-2}, it is evident that both the light curves are very well matched. The mean of the difference light curve (bottom panel of Fig. \ref{figure-2}) was found to be 0.016 $\pm$ 0.003 mag. This is due to the systematically fainter magnitudes from {\it Photutils} as the derived magnitudes were not corrected for saturation effects. The advantage of {\it Curvit} over {\it Photutils} is that {\it Curvit} works on the calibrated events list, while {\it Photutils} works on the images. Therefore, for the rest of the sample sources, the {\it Curvit} package  \citep{2021JApA...42...25J} was used to generate the light curves. In each source, the individual orbit-wise events list were combined using {\it Curvit} package and the combined events list was used for generating the light curves. In addtion to background subtraction, the light curves were also subjected to aperture and saturation correction. The brightness of the blazars derived over a circular aperture of 12 sub-pixel radii using {\it Curvit} was then converted to the total brightness using Table 11 of \cite{2020AJ....159..158T}. The brightness of all the blazars thus obtained is given in the Appendix.

\section{Analysis}
\label{sec:analysis}

\subsection{Flux variability}
The intrinsic amplitude of flux variability, which is basically the observed variance of the light curve after the removal of the measurement errors, was calculated to characterize variability. The intrinsic amplitude of flux variability, $\sigma_m$ (mag), is defined as \citep{2010ApJ...716L..31A},
\begin{equation}
%\hspace{3cm}
{\sigma_m = \sqrt {(\Sigma^2 - \epsilon^2)}}
\end{equation}
Here,  $\Sigma^2$ is the observed variance and is given by
\begin{equation}
%\hspace{2.5cm}
\Sigma = \sqrt{\frac{1}{N-1} \sum_{i=1}^{N} (m_i - \langle m \rangle)^2}
\end{equation}
where, N is the number of orbits, $\langle m \rangle$ is the weighted mean of the $m_i$ measurements with  errors $\epsilon_i$. $\epsilon$ is the contribution of measurement errors to the variance and is defined as
\begin{equation}
%\hspace{3cm}
\epsilon^2 = \frac{1}{N}\sum_{i=1}^{N} \epsilon_i^2
\end{equation}
A source was considered variable if $\Sigma^2$ $>$ $\epsilon^2$, else, \\
$\sigma_m$ = 0. 

\subsection{Spectral variability}
\label{sec: spectral variability procedure}

In order to study the various spectral trends found in blazars, different colour-index combinations were obtained and plotted against the magnitude, popularly known as the color-magnitude diagram (CMD). In general, in the optical band most BL Lacs exhibit a bluer-when-brighter (BWB) trend \citep[]{vagnetti2003spectral,li2024optical}, while FSRQs display a redder-when-brighter (RWB) trend \citep[]{gu2006multi, negi2022optical}. Color variations on hour time scales for five sources having more than four simultaneous FUV and NUV photometric observations were studied. To study the color trends quantitatively, the CMDs for all combinations were fitted with an unweighted linear least-squares fit of the form, color index $(CI) = m\times M + c$, where, $M$ is the magnitude, and $m$ and $c$ are slope and intercept, respectively. A positive value of the slope indicates a BWB trend, while a negative value hints toward the presence of a RWB trend. For the trend to be significant at the 99\% significance level, the correlation coefficient ($R$) is to be more than 0.5 \citep{2021A&A...654A..38P}. In addition to the unweighted least-squares fit to the points on the CMD, a weighted linear least-squares fit was also carried out by considering the errors in both the colors and the magnitudes. For this, a Bayesian linear regression with the LINMIX$\_$ERR method \citep{kelly2007} was used. This method is effective in handling errors in both the color and magnitude measurements as well as correlations between the errors. In addition to LINMIX$\_$ERR, the Bivariate Correlated Errors and Intrinsic Scatter (BCES; \citealt{akritas1996}) was also used, and the results are consistent with each other. For all the further discussions on spectral variability, the results from the Bayesian analysis are used. 

\begin{deluxetable}{lclrc}
\tablecaption{Results of variability analysis for sources showing variability. Here, N is the number of orbits and $\sigma_m$ is the amplitude of flux variability in magnitude.\label{Table-4}}
\tabletypesize{\footnotesize}
\tablehead{
Name & \colhead{Date} & \colhead{Filter} & \colhead{N}  & \colhead{$\sigma_m$} \\
& \colhead{(dd-mm-yyyy)}  &    &          &  (mag)      
}
\startdata
PKS 0208$-$512 & 30-10-2016    & F148W  & 4  & 0.06 \\
1ES 0229+200   & 01-10-2017    & N245M  & 3  & 0.03 \\
               & 09-12-2017    & F154W  & 4  & 0.12 \\
               & 08-01-2018    & F154W  & 4  & 0.04 \\
               &               & N245M  & 4  & 0.05 \\
OJ 287         & 10-04-2017    & F169M  & 11 & 0.03 \\
               &               & N245M  & 7  & 0.01 \\
               & 18-04-2018    & F169M  & 24 & 0.03 \\
               & 18-05-2020    & F148W  & 6  & 0.01 \\
               &               & F154W  & 8  & 0.04 \\
               &               & F172M  & 10 & 0.04 \\
               &               & F148Wa  & 6  & 0.03 \\
1ES 1101$-$232 & 30-12-2016    & F154W  & 32 & 0.04 \\
               &               & N263M  & 32 & 0.03 \\
1ES 1218+304   & 21-05-2016    & F148W  & 13 & 0.01 \\
H 1426+428     & 05-03-2018    & F148W  & 8  & 0.03 \\
               &               & F154W  & 5  & 0.02 \\
               &               & N242W  & 4  & 0.03 \\
PKS 1510$-$089 & 16-03-2018    & F172M  & 11 & 0.02 \\
               &               & N219M  & 11 & 0.12 \\
PKS 2155$-$304 & 21-05-2018    & F154W  & 4  & 0.02 \\
               & 16-09-2019    & F154W  & 3  & 0.02 \\
1ES 2344+514   & 06-06-2017    & F172M  & 3  & 0.98 \\
               & 09-07-2017    & F172M  & 3  & 0.06 \\
               &               & N245M  & 3  & 0.03 \\
               & 07-09-2017    & F172M  & 3  & 0.02 \\
               & 07-12-2017    & F172M  & 2  & 0.24 \\ 
\enddata
\end{deluxetable}

\begin{table*}
\centering
\caption{Results of unweighted linear least-squares fit to the colour magnitude diagram. Here, column 1 gives the name of the sources, column 2 gives the OBSID, column 3 gives the number of orbits used in the analysis, column 4 gives the FUV and NUV filters used to get the colour and brightness, column 5 gives the slope and the associated error, column 6 gives the intercept and the associated error, column 7 gives the linear correlation coefficient (R) and column 8 gives the probability of no correlation (P). The entries with 0 in P (column 6) indicates that the values are lesser than $1 \times 10^{-2}$.}
\begin{tabular}{l@{\hspace{0.3cm}}c@{\hspace{0.2cm}}c@{\hspace{0.2cm}}c@{\hspace{0.25cm}}c@{\hspace{0.25cm}}l@{\hspace{0.25cm}}c@{\hspace{0.25cm}}c@{\hspace{0.25cm}}} 
\hline
\hline
Name  &  OBSID & No. of  & Color/Mag & Slope & Intercept & R & P  \\
&    &   orbits & & & & & \\
\hline
OJ 287 & T01-163T01-9000001152 & 07 & (F169M - N245M)/F169M   & 1.04 $\pm$ 0.20 & $-$16.53 $\pm$ 3.34 &  0.92  & 0.00  \\
1ES 1101$-$232 & G06-086T02-9000000936 &  32 & (F154W - N263M)/F154W  & 1.01 $\pm$ 0.12 & $-$18.13 $\pm$ 2.20  & 0.84 &  0.00  \\
1ES 1218+304 & G05-211T01-9000000464 & 13 & (F148W - N245M)/F148W & 1.16 $\pm$ 0.14 & $-$20.23 $\pm$ 2.46 &  0.93 &  0.00 \\ 
H 1426+428 & A04-094T01-9000001942 &  07 & (F148W - N245M)/F148W & 0.66 $\pm$ 0.38 &  $-$11.79 $\pm$ 7.16   &  0.61 &  0.15 \\
PKS 1510$-$089 & A04-101T01-9000001984 & 11 & (F172M - N219M)/F172M &  1.85 $\pm$ 0.58 & $-$32.19 $\pm$ 10.03 &  0.73 &  0.01  \\       
\hline
\end{tabular}
\label{Table-5}
\end{table*}

\begin{table*}
\centering
\caption{Results of fit to the points in the CMD using BCES and Bayesian methods. The columns have the meaning as given in Table \ref{Table-5}.}
\begin{tabular}{l@{\hspace{0.15cm}}c@{\hspace{0.1cm}}c@{\hspace{0.05cm}}c@{\hspace{0.15cm}}c@{\hspace{0.15cm}}c@{\hspace{0.15cm}}c@{\hspace{0.15cm}}c@{\hspace{0.15cm}}} 
\hline
\hline
Name  &  OBSID & No. of  & Color/Mag & \multicolumn{2}{c}{BCES} & \multicolumn{2}{c}{Bayesian}\\
      &        &      orbits         &           &   Slope    &  Intercept     &  Slope   &  Intercept   \\ 
\hline
OJ 287         & T01-163T01-9000001152 & 07  & (F169M - N245M)/F169M   & 1.08  $\pm$ 0.18 &  $-$17.18 $\pm$ 3.01 & 1.15 $\pm$ 1.02  & $-$18.31 $\pm$ 16.74  \\
1ES 1101$-$232 & G06-086T02-9000000936 & 32 & (F154W - N263M)/F154W  & 1.08 $\pm$ 0.17 & $-$19.31 $\pm$ 3.19 & 1.29 $\pm$ 0.18  &  $-$23.35 $\pm$ 3.37   \\
1ES 1218+304   & G05-211T01-9000000464 & 13 & (F148W - N245M)/F148W & 1.04 $\pm$ 0.11 & $-$18.14 $\pm$ 2.06 & 1.14  $\pm$ 0.54 &  $-$19.84  $\pm$ 9.70\\ 
H 1426+428     & A04-094T01-9000001942 & 07  & (F148W - N245M)/F148W & 0.60 $\pm$ 0.32 & $-$10.78 $\pm$ 5.97 & 1.07 $\pm$ 2.08 &  $-$19.50 $\pm$ 38.80  \\
PKS 1510$-$089 & A04-101T01-9000001984 & 11 & (F172M - N219M)/F172M  & 2.75 $\pm$ 0.67 & $-$47.89 $\pm$ 11.66 & 2.43 $\pm$ 1.74  &  $-$42.29 $\pm$ 30.10   \\ 
\hline
\end{tabular}
\label{Table-6}
\end{table*}

\begin{figure*}
\centering
\includegraphics[scale=0.55]{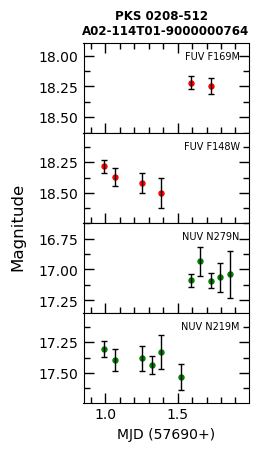}
\includegraphics[scale=0.55]{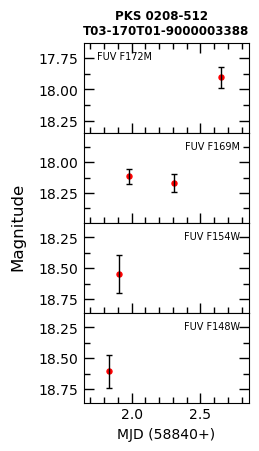}
\includegraphics[scale=0.55]{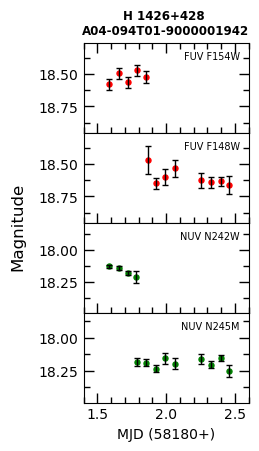}
\includegraphics[scale=0.55]{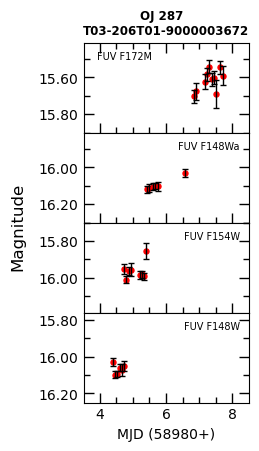}

\includegraphics[scale=0.55]{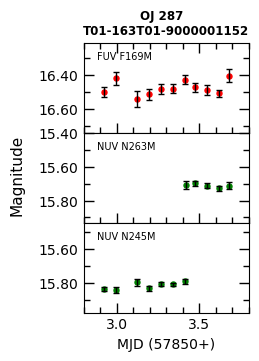}
\includegraphics[scale=0.55]{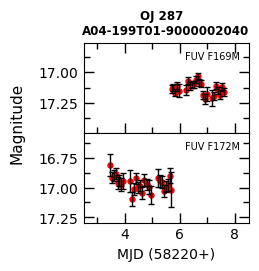}
\includegraphics[scale=0.55]{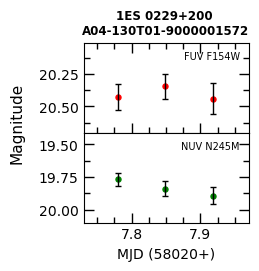}
\includegraphics[scale=0.55]{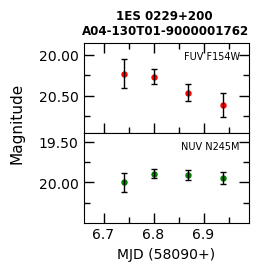}

\includegraphics[scale=0.55]{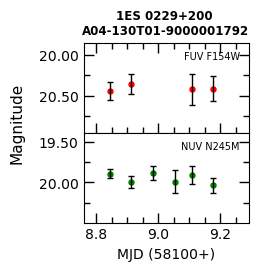}
\includegraphics[scale=0.55]{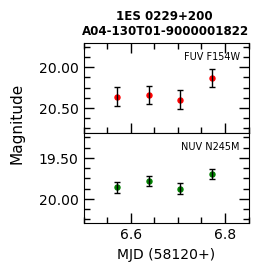}
\includegraphics[scale=0.55]{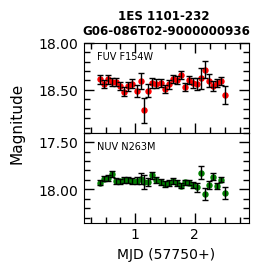}
\includegraphics[scale=0.55]{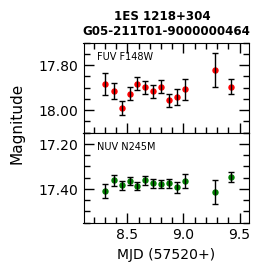}

\includegraphics[scale=0.55]{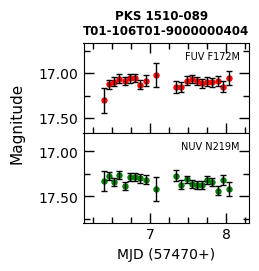}
\includegraphics[scale=0.55]{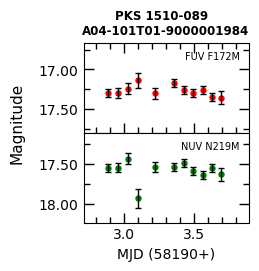}
\includegraphics[scale=0.55]{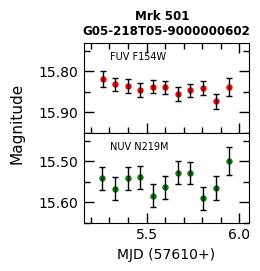}
\includegraphics[scale=0.55]{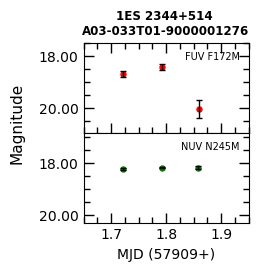}

\includegraphics[scale=0.55]{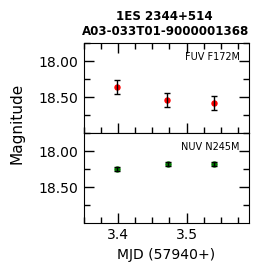}
\includegraphics[scale=0.55]{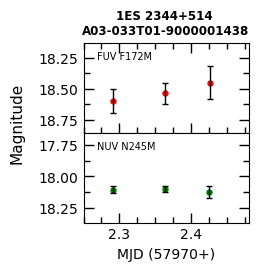}
\includegraphics[scale=0.55]{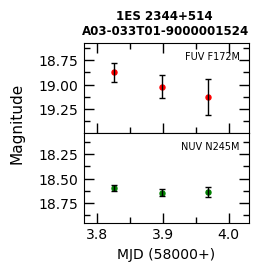}
\includegraphics[scale=0.55]{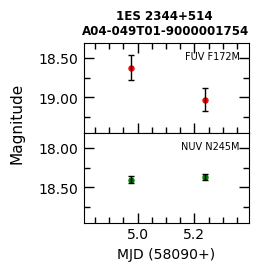}

\caption{FUV and NUV light curves (represented by red and green color circles, respectively) of sources. The observation ID and source name are labeled in each plot. The name of the filters used for the light curves are given in each panel.}
\label{figure-3}
\end{figure*}

\begin{figure*}
\centering
\includegraphics[scale=0.65]{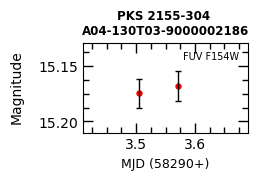}
\includegraphics[scale=0.65]{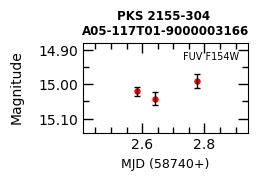}
\includegraphics[scale=0.65]{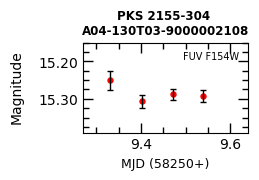}
\includegraphics[scale=0.65]{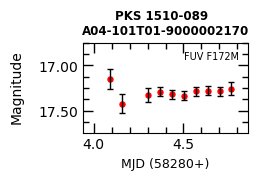}

\flushleft
{
\hspace{0.3cm}
\includegraphics[scale=0.65]{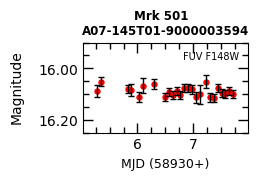}
\includegraphics[scale=0.65]{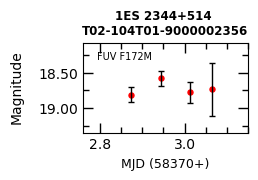}
\includegraphics[scale=0.65]{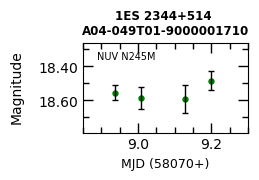}
}
\caption{FUV and NUV light curves of sources. Here, red dots denote FUV light curves and green dots denote NUV light curves. The respective source name and observation ID are mentioned on top of each plot. Also, the names of the filters used for the light curves are given in each panel.}
\label{figure-4}
\end{figure*}

\begin{figure*}
\hbox{
\hspace{0.3cm}
\includegraphics[scale=0.45]{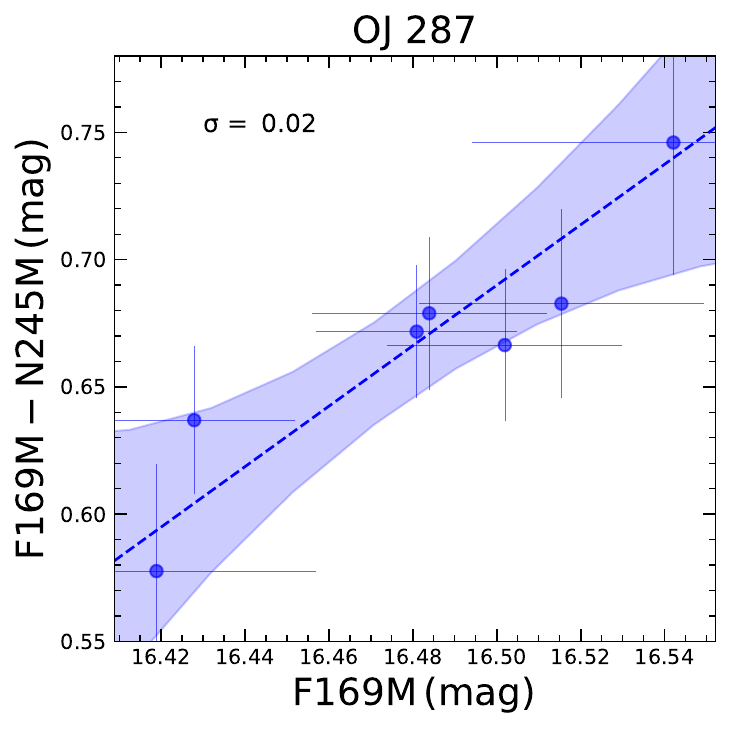}
\includegraphics[scale=0.45]{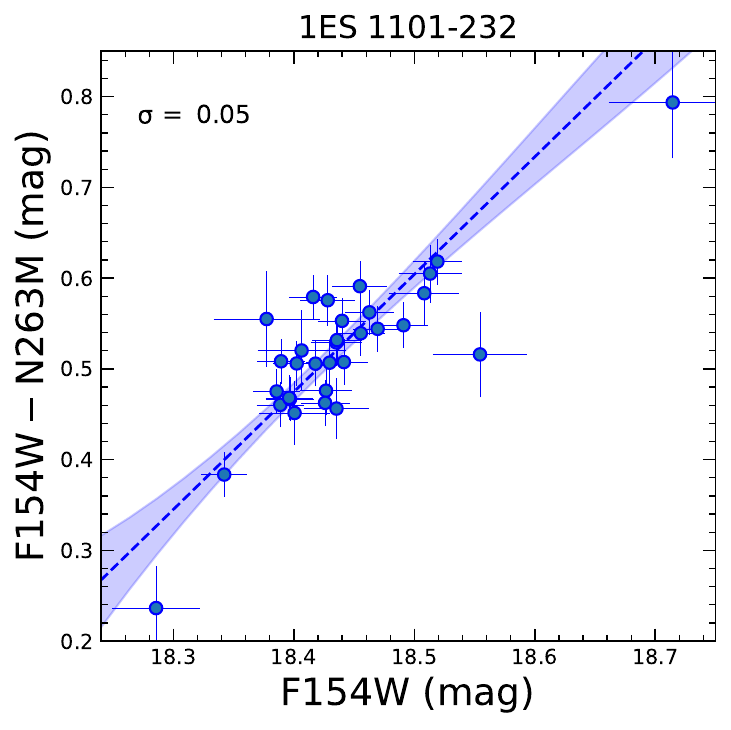}
\includegraphics[scale=0.45]{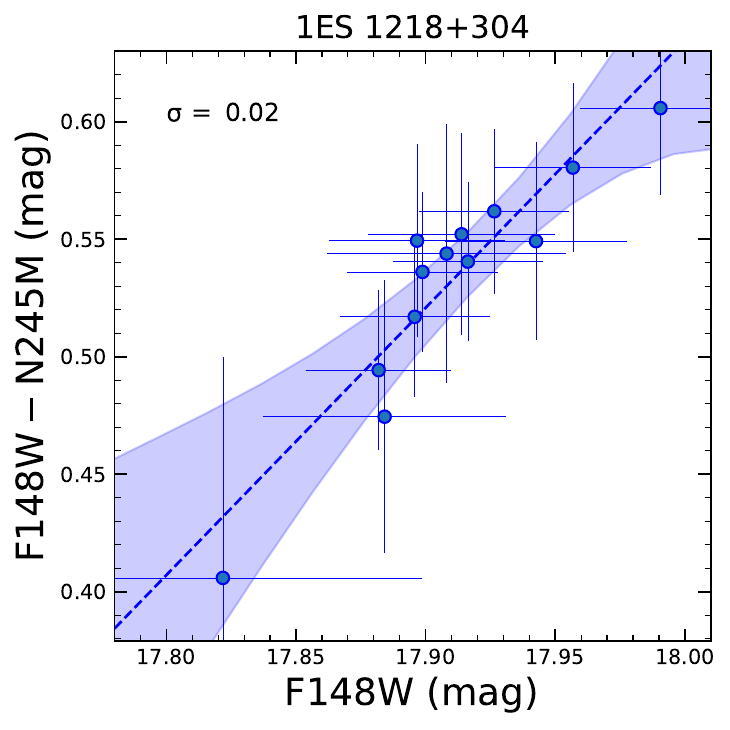}
}

\hbox{
\hspace{0.3cm}
\includegraphics[scale=0.46]{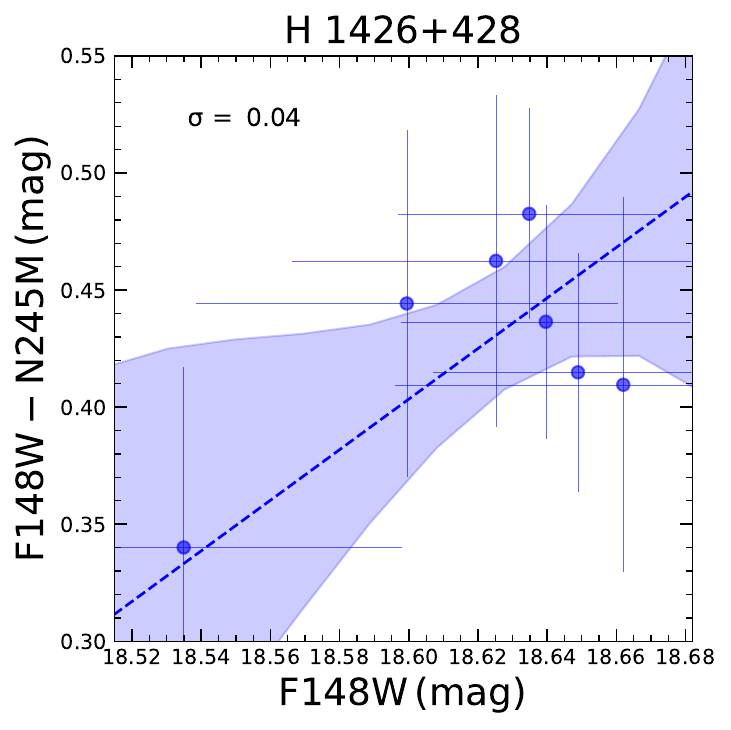}	
\includegraphics[scale=0.46]{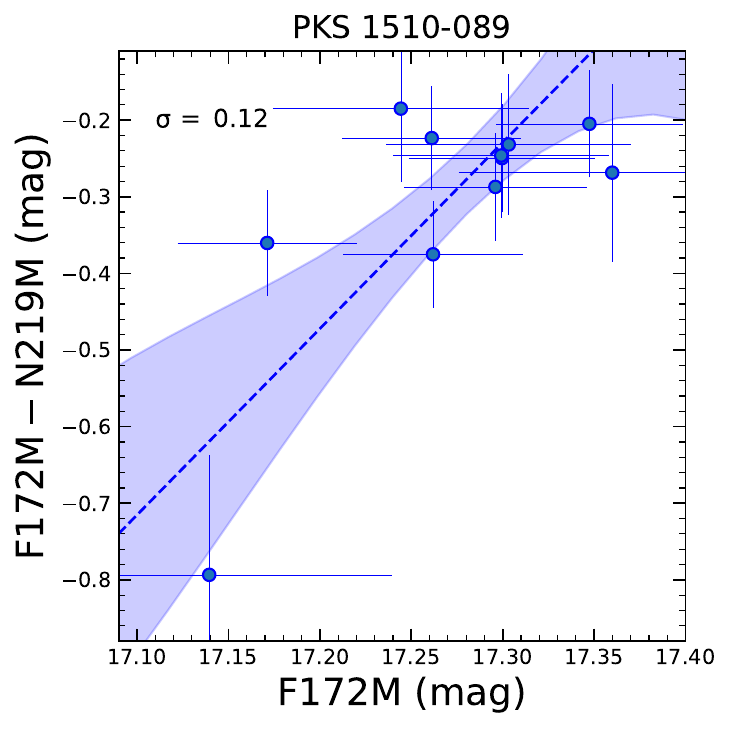}
}

\caption{CMDs for the sources that are variable and having more than four simultaneous FUV and NUV photometric data points. The best fit line from Bayesian analysis using LINMIX$\_$ERR is shown by the dashed blue line and the uncertainty range on the best fit is shown by the blue shaded region. The intrinsic scatter ($\sigma$) is given in the respective plots. The names of the sources are given on the top of each panel (see also Table \ref{Table-5} and \ref{Table-6}).}
\label{figure-5}
\end{figure*}

\section{Results}
\label{sec:results}

\subsection{Flux variability}
All the ten blazar fields used in this work are less crowded. As the goal of the present study is to characterize the UV variability of blazars on hour-like time scales, aperture photometry was carried out following the procedures outlined in Section \ref{sec:Observations and Data Reduction}. Of all the sources analyzed in this work, 9 sources were found to show flux variations, except that of Mrk 501, which was found to be non-variable. The results of the variability analysis for sources showing variability are given in Table \ref{Table-4}, and the light curves generated for all the sample sources are given in Figs. \ref{figure-3} and \ref{figure-4}. Further, details of the flux variability of each of the sources are given in Section \ref{sec:source description}.

\subsection{Spectral variability }
Flux variations in blazars are accompanied by spectral variations. The color changes in blazars on diverse time scales directly related to the spectrum of the blazar are well-studied by many authors in optical bands \citep{2004A&A...421..103V, 2009MNRAS.399.1357S, 2012ApJ...756...13B, 2018Galax...6....2B, 2019MNRAS.484.5633G, 2023ApJ...946..109A}, but genuine color behavior at different time scales is still one of the most puzzling issues in blazar physics. Moreover, color changes on hourly time scales in the UV bands are rarely studied in blazars, though few studies exist for Seyfert-type AGN \citep{2022MNRAS.511L..13C}. This is the first search for dominant color trends in blazars on the intraday time scale in the UV regime. A better understanding of dominant emission mechanisms in the source can be understood through the studies on spectral variations. The UV spectral variations can be contributed by both the accretion disk and the Doppler-boosted relativistic jets. In order to understand this, CMDs were generated for the sources. For this, the five sources which are variable and also having more than four simultaneous NUV and FUV measurements were only considered. The CMDs were then analyzed following the methodology described in Section \ref{sec: spectral variability procedure}. The fits to the CMDs of the sources are given in Fig. \ref{figure-5}, and the results of the fit are given in Tables \ref{Table-5} and \ref{Table-6}. BWB trend was observed for all the sources analysed for spectral variability. Studies in the literature \citep{2002A&A...390..407V, 2017Natur.552..374R} suggest that a BWB trend in blazars is due to intrinsic energetic processes in the jet.

\subsection{Notes on individual sources}
\label{sec:source description}

\subsubsection{PKS 0208$-$512}  
PKS 0208$-$512 is a FSRQ of LSP type at a redshift of $z$ = 1.003 \citep{2008ApJS..175...97H}. 
It has been detected by EGRET on board the {\it Compton Gamma Ray Observatory} \citep{1999ApJS..123...79H} and {\it Fermi} Gamma-Ray Space Telescope \citep{2020ApJS..247...33A}. It has been studied for 
optical and infrared flux variations on a day-like time scale \citep{2012ApJ...756...13B}.  
Between August 2008 and September 2011, 
the source underwent three optical and infrared bursts from monitoring observation on day-like 
time scales. Of the three optical infrared
outbursts, only two outbursts and flares at GeV energies were observed. A RWB
spectral variation was noticed in the optical and near-infrared bands \citep{2013ApJ...763L..11C}.
It was found to show intra-night optical variability \citep{2002A&A...390..431R}.
Variations in the UV band are reported for the first time.

\subsubsection{1ES 0229+200}  
This is a very high energy $\gamma$-ray source \citep{2007A&A...475L...9A} and also detected in the GeV energy range by the {\it Fermi} Gamma-Ray Space Telescope \citep{2015ApJ...810...14A}. It was found to show optical flux variations within a night \citep{2012MNRAS.424.2625B} and on day-like time scales \citep{2020MNRAS.496.1430P} and VHE variations on month-like time scales \citep{2015ICRC...34..762C}. It was also studied for X-ray flux variations in the 3$-$79 keV band using observations from {\it NuSTAR} \citep{2017ApJ...841..123P}. It has not been studied for UV variations before. It was found to be variable on all the four epochs it was observed.

\subsubsection{OJ 287}
This source, believed to host a supermassive binary black hole \citep{1988ApJ...325..628S}, has been 
extensively studied  for flux variations across wavelengths on a range of time scales, including UV \citep{2012MNRAS.424.2625B,2022MNRAS.515.2778H}, optical \citep{1978ApJS...38..267O} and
GeV $\gamma$-rays \citep{2011ApJ...726L..13A}. In addition to flux variations, the source has also been studied for optical polarization variation \citep{2017ApJ...835..275R}. This source was observed for three epochs simultaneously in both FUV and NUV. The source was found to show variations on all three epochs. In addition to flux variations, the source was also found to show spectral variations in one of the epochs with a BWB trend.

\subsubsection{1ES 1101$-$232}
This is a very high energy BL Lac \citep{2007A&A...470..475A}. It was found to show variations in the 3$-$79 keV band in X-rays \citep{2018ApJ...859...49P}, optical flux \citep{2002A&A...390..431R} and polarization microvariability. This source has no previous record on its UV flux variations. This source was observed by UVIT on 30 December 2016, with two different filters (F154W and N263M). The source was found to show flux variations on hour-like time scale in both FUV and NUV bands. In addition to flux variability, the source was also found to show spectral variability with a BWB pattern.

\subsubsection{1ES 1218+304}
It is an HSP blazar at a redshift $z$ = 0.182 \citep{2022MNRAS.511.5611O}. It is also detected in GeV $\gamma$-rays by {\it Fermi} \citep{2012ApJS..199...31N} and in very high 
energy $\gamma$-rays from \textit{MAGIC} \citep{2006ApJ...642L.119A} and from \textit{VERITAS} \citep{2009ApJ...695.1370A} observations. It has been studied for optical flux variations both in long-term \citep{2022MNRAS.510.1791N} and within a night \citep{2011MNRAS.416..101G}. It is highly polarized in the optical with a polarization degree of 6.83 $\pm$ 0.70\%
\citep{1994ApJ...428..130J}.  It  has been found to be variable in the X-ray band in the 0.3$-$10 keV energy range from \textit{XMM-Netwon} observations \citep{2022ApJ...939...80D}, in the hard X-ray band from \textit{NuSTAR} observations \citep{2018ApJ...859...49P} and in the high energy $\gamma$-ray band  from \textit{VERITAS} \citep{2010ApJ...709L.163A}.  From \textit{Swift}/UVOT observations 1ES 1218+304 was found to show long term variations in UV \citep{2022ApJ...931...83M}. However, it has not been studied for UV variations on hour-like time scales before. The source was observed on one epoch. On this epoch, the source was found to show flux variations as well as a BWB spectral variability.

\subsubsection{H 1426+428}
It is an extreme HSP blazar \citep{2001A&A...371..512C} and was detected in TeV $\gamma$-rays from observations with the Whipple telescope \citep{2002ApJ...571..753H} and
from \textit{HEGRA} observations \citep{2002A&A...384L..23A}. In the radio band, it has a compact core surrounded by a halo \citep{2004ApJ...613..752G}. From \textit{VLBA} observations at 8 GHz, \cite{2008ApJ...678...64P} found the source to have a parsec scale radio structure consisting of a core and a single jet component. It was observed to show optical flux variations within a night \citep{2011MNRAS.416..101G}. It was found to show long-term flux and color variations in the optical \citep{2012MNRAS.425.3002G, 2022MNRAS.510.1791N} as well as flux variation in X-rays \citep{2022ApJ...939...80D}. It is polarized in the optical and also show polarization variations \citep{2016A&A...596A..78H}. This source has been studied for UV flux variations for the first time. The source was found to show both flux and spectral variations (BWB trend) on the single epoch it was observed.

\subsubsection{PKS 1510$-$089}
It is an FSRQ at $z$ = 0.360 \citep{1990PASP..102.1235T} and powered by a black hole 
of mass 5.7$^{+0.62}_{0.58}$ $\times$ 10$^7$ M$_{\odot}$ \citep{2020A&A...642A..59R}. 
It is known to emit high energy
$\gamma$-rays in GeV band and also in TeV from HESS \citep{2013A&A...554A.107H} and
MAGIC \citep{2018A&A...619A.159M}. Quasi-periodic oscillation (QPO) was detected in the GeV light curve of the source
\citep{2022MNRAS.510.3641R}. PKS 1510$-$089 has been studied for correlation between
flux variations between different wavelengths. The flux variations in the $\gamma$-ray and optical/IR
bands were found to precede the variations in the radio band \citep{2023ApJ...953...47Y}. While
\cite{2016MNRAS.456..171R} found the long-term optical and $\gamma$-ray flux variations to
be correlated with zero lag, \cite{2020MNRAS.498.5128R} found complex flux variability behavior
between optical and $\gamma$-ray bands. In the broad band SED, optical/UV emission was found to be 
dominated by emission from the accretion disk \citep{barnacka2014pks}. It was known to show short time scale flux variations in the optical \citep{2011MNRAS.416..101G,1999A&AS..135..477R,2002A&A...390..431R} and $\gamma$-rays \citep{2013ApJ...766L..11S}. It is polarized
in the optical and also shows polarization variations \citep{2005A&A...442...97A}. From
observations carried out with IUE, the source was found to show variability in the UV 
continuum \citep{1988ESASP.281b.261O}. This source has not been studied before for short-term variations in
the UV. This source was detected in three epochs. It was found to show flux and spectral variability in one of the epochs.

\subsubsection{Mrk 501}
This nearby source ($z$ = 0.034) is a strong X-ray and TeV $\gamma$-ray source 
\citep{1998ApJ...492L..17P, 1999A&A...350...17D} with a black hole of mass
10$^{8.93 \pm 0.21}$ M$_{\odot}$ \citep{2002ApJ...569L..35F}. In the radio band, at the 
parsec scale, the source is known to have a relativistically boosted one-sided jet
\citep{2002ApJ...579L..67E}. It is known to be variable in X-rays and
TeV $\gamma$-rays \citep{2006ApJ...646...61G}. In the optical, rapid variations
on the time scale of minutes have been observed \citep{2019PASP..131g4102Z}. In addition to flux
variations, optical color variations have also been observed. This source was observed for two epochs to know their UV variability nature on hour-like time scales. On both epochs, it was found that the
source does not show any variability.

\subsubsection{PKS 2155$-$304}
This HSP blazar situated at $z$ = 0.116 \citep{1993ApJ...411L..63F}  has been 
extensively studied across wavelengths. It is variable in the 
optical \citep{2010ApJ...718..279G,1999A&AS..135..477R}, 
X-rays \citep{2021ApJ...909..103Z}, $\gamma$-rays \citep{2010A&A...520A..23R}, 
UV from observations with the IUE \citep{1993ApJ...411..614U} and in the extreme UV from
the Extreme Ultraviolet Explorer observations \citep{2001ApJ...549..938M}. QPO has
been detected in the optical \citep{2014ApJ...793L...1S}, X-ray \citep{2009A&A...506L..17L} and
GeV $\gamma$-rays \citep{ren2023}. It is polarized
in the optical and also shows polarization variations \citep{2020MNRAS.495.2162P}. This source was observed for three epochs in the FUV band, and on all the occasions, it was found to be variable.

\subsubsection{1ES 2344+514}

The source 1ES 2344+514  at a redshift of $z$ = 0.044 \citep{1996ApJS..104..251P} is an X-ray source detected in the Einstein Slew Survey \citep{1992ApJS...80..257E} and is also a VHE $\gamma$-ray emitter \citep{1998ApJ...501..616C}. It is also detected by VERITAS \citep{Acciari2011} and MAGIC \citep{MAGICCollaboration2020}.
It was found to have short time scale variation in X-rays \citep{Giommi2000} and optical \citep{Pandey2020}. In the optical, it was also found to show variations on day-like time scales \citep{Dai2001,Gaur2012,Cai2022}. It is also variable in $\gamma$-rays \citep{Grube2008}. On long-time scales, it is found to be variable in UV based on observations with the Swift/UVOT acquired over a period of three years from 2019 to 2021 \citep{Abe2024}. This source has been observed for seven epochs to know its UV variability nature on hour-like time scale. The source was found to show flux variability on four epochs.

\section{Discussion and Conclusions}
\label{sec:Discussion}

Flux variability in blazars in UV is found to be extremely useful for getting insights into the inner region close to the SMBH, radiation mechanisms taking place in the vicinity of SMBH, and the relativistic jets. However, because intranight observations using space-based UV telescopes are time-expensive, the nature of UV variability in blazars is not well-studied in the literature. Recently, several studies have been conducted on large samples of optically selected quasars using their UV data obtained with \textit{GALEX} \citep{2016ApJ...830..104P, 2021ApJ...919...40P} and covering a day or longer timescales. Here, the first study was performed to explore the properties of blazars in UV on the shorter, hour-like time scales. For this purpose, a sample of 10 blazars was concentrated, comprising of 2 FSRQs and 8 BL Lacs objects that were observed using UVIT payload onboard \textit{AstroSat}.
\\[6pt]
In blazars, the UV emission obtained by the instrument is widely believed to be non-thermal synchrotron emission along the jet axis. Doppler factor $\delta$ is defined as $\delta$ = $1 / {\Gamma(1-\beta \cos\theta_{obs})}$, where $\Gamma$ is the bulk Lorentz factor, $\beta$ is the speed of emission region in the units of speed of light, and $\theta_{obs}$ is the angle of the jet with respect to the line of sight to the observer. In blazars, the observer's line of sight 
 is closely aligned with the orientation of the relativist jet \citep{1995PASP..107..803U}, and thus, the jet emission undergoes Doppler boosting owing to a higher apparent Lorentz factor, and thus, the jet emission overpowers the thermal radiation from the accretion disc. This is followed by a series of effects, such as amplified flux and blueshifted emitted frequencies, which causes the shortening of their apparent variability time scales \citep{1995PASP..107..803U}. Because of the above reasons, blazar variability is an excellent tool for shedding light on the structure of jets and understanding the emission mechanisms taking place in them.
\\[6pt]
Various theoretical models used to explain blazar variability are broadly divided into extrinsic and intrinsic mechanisms. The probable intrinsic mechanisms include shocks traveling down the jet or magnetic irregularities or accretion disc instabilities \citep[][and references therein]{1979ApJ...232...34B,2008Natur.452..966M, 2015ApJ...805...91M}. Whereas the extrinsic effects involve interstellar scintillation as well as variations in the viewing angle leading to changes in the Doppler factor, or gravitational microlensing \citep[][and references therein]{1987A&A...171...49S}. During the low state in blazars, UV variability can be attributed to the instabilities or hot spots on the accretion disk \citep{1993A&A...271..216C}. However, various multi-wavelength studies of these sources have demonstrated that the electromagnetic emission from all these blazars is jet-dominated \citep{barnacka2014pks, sahakyan2020broad, prince2021multiwavelength, kushwaha2021astrosat,  prince2021comprehensive, diwan2023multiwavelength, goswami2024variety}. Therefore, it is tempting to postulate that the observed variable UV emission from them could be largely contributed by their Doppler boosted relativistic jet. To further investigate this claim, we analysed the colour-magnitude relationship of the sample sources. 
\\[6pt]
Blazars, a subclass of AGN, exhibit intriguing color or spectral index variations that play a crucial role in unraveling their underlying emission mechanisms. These enigmatic cosmic objects exhibit various behaviors in their CMDs. CMDs of blazars have shown to exhibit three distinct trends namely, RWB, BWB and achromatic behaviours. Generally, most studies indicate that BWB chromatism is the dominant behavior in most of the BL Lac objects, while RWB chromatism is typical of FSRQs \citep[][and references therein]{2003ApJ...590..123V, 2022ApJ...933...42A, 2023ApJS..265...51A}. However, the color-magnitude correlation in blazars still remains a topic of debate in literature. The BWB trend is often explained by a single-component synchrotron model where an influx of freshly injected electrons with a harder energy distribution than the previous cooler electrons leads to increased flux and a bluer spectrum. Also, electrons accelerated at the shock front have higher energies, they gradually loose energy due to radiative cooling, thereby showing more variations at the higher energy band \citep[]{1998A&A...333..452K,mastichiadis2002models}. Existence of an energy stratified jet also finds support from the energy dependent polarization observed recently in HSP blazars such as Mrk 421 \citep{kim2024magnetic} and 1ES 0229+200 \citep{ehlert2023x}. Furthermore, this trend can also be elucidated by a two-component emission model. In this scenario, a stable component with a constant spectral index ($\alpha_{constant}$) coexists with a variable component characterized by a flatter slope ($\alpha_{1}$). As the variable component outshines the stable one, chromatic behaviors emerge, influencing short-term behavior (dominated by pronounced chromatic components) and longer-term variations (attributed to a mildly chromatic component). In all the sources studied for spectral variations, a BWB behavior was found, also suggesting that the variable flux seen in the sample sources could be attributed to synchrotron emission from the jet, potentially signaling an enhancement in particle acceleration efficiency. One of the FSRQs, namely PKS 1510$-$089, also displayed the BWB trend, thus suggesting that the jets of BL Lacs and FSRQs could be fundamentally the same. 
\\[6pt]
Detailed spectral measurements of a handful of blazar jets based on examination of flux densities of the knots in the kiloparsec-scale jets hint towards the fact that a significant portion of their synchrotron UV radiation might come from relativistic particles different from those responsible for optical/IR emission. The study displays an increase toward the UV and a smooth transition to X-ray data points, thus revealing a clear spectral upturn in the optical/UV region of their spectral energy distributions \citep{2006ApJ...648..910U}. The presence of high polarization in the UV emission, as found by \citet{2013ApJ...773..186C}, supports the origin of this high energy component. The above findings have also been confirmed by \citet{2022MNRAS.511L..13C} by studying the optical and UV emissions (rest-frame) from blazars.
\\[6pt]
The scarcity of information on the UV variability in blazars is due to the expensive nature of carrying out intranight monitoring campaigns using space-based UV telescopes \citep{2018ApJ...857..141S}. To tackle this challenge, recent efforts have leveraged UV data collected by \textit{GALEX} for large samples of AGN. However, these studies mainly focused on day-long or longer time scales monitoring of the optically selected quasars \citep{2016ApJ...830..104P}, thereby examining the accretion disc rather than the relativistic jet of these sources. To delve into the rapid UV variability of blazars, UV flux variability characteristics on hourly time scales of 10 blazars observed using UVIT on board \textit{AstroSat} in FUV and NUV filters were studied. UV flux variations on time scales of hours was found in 9 blazars. Also, the color-magnitude relationship was analyzed for the sample of sources that are found to be variable and having more than 4 simultaneous observations in FUV and NUV filters namely OJ 287, 1ES 1101$-$232, 1ES 1218+304, H 1426+428 and PKS 1510-089. Of these the first four sources belong to the BL Lac category, while PKS 1510$-$089 is a FSRQ. Thus, both FSRQs and BL Lacs in the sample were found to display a dominant BWB chromatism in the UV region. The above findings indicate the dominance of the synchrotron emission from the Doppler-boosted jet for this sample of blazars in UV band.

\section*{Acknowledgements}
The authors extend their sincere gratitude to the anonymous reviewer for the useful comments and suggestions on the manuscript. This research made use of Photutils, an Astropy package for detection and photometry of astronomical sources \citep{larry_bradley_2023_1035865}. This publication uses the data from the \textit{AstroSat} mission of the Indian Space Research Organisation (ISRO), archived at the Indian Space Science Data Centre (ISSDC). This publication uses UVIT data processed by the payload operations centre at Indian Institute of Astrophysics (IIA), Bangalore. The UVIT is built in collaboration between IIA, Inter University Center for Astronomy and Astrophysics (IUCAA), Tata Institute of Fundamental Research (TIFR), ISRO and Canadian Space Agency (CSA). One of the authors (SBG) thanks the IUCAA, Pune, India for the Visiting Associateship.

\bibliography{ref}
\bibliographystyle{aasjournal}
\newpage
\appendix
\label{sec:appendix}

\section{Brightness measurements of the source PKS 0208$-$512}
\begin{deluxetable*}{ccccc}[htbp]
\tablehead{\colhead{Date}  & \colhead{Filter} &  \colhead{UTstart} & \colhead{UTend} & \colhead{Mag}}
\startdata
30-10-2016  & F148W &  57690.991247  &  57690.998305  &  18.28  $\pm$  0.05  \\
            &       &  57691.063816  &  57691.067698  &  18.37  $\pm$  0.07  \\
            &       &  57691.249972  &  57691.253182  &  18.42  $\pm$  0.08  \\
            &       &  57691.385321  &  57691.386868  &  18.50  $\pm$  0.12  \\
            & F169M &  57691.588337  &  57691.600612  &  18.22  $\pm$  0.05  \\
            &       &  57691.723686  &  57691.731484  &  18.25  $\pm$  0.07  \\
            & N219M &  57690.991175  &  57690.998355  &  17.31  $\pm$  0.07  \\
            &       &  57691.063744  &  57691.067748  &  17.39  $\pm$  0.09  \\
            &       &  57691.249923  &  57691.253181  &  17.38  $\pm$  0.10  \\
            &       &  57691.317597  &  57691.324360  &  17.44  $\pm$  0.07  \\
            &       &  57691.385272  &  57691.386868  &  17.33  $\pm$  0.14  \\
            &       &  57691.520614  &  57691.524076  &  17.53  $\pm$  0.10  \\
            & N279N &  57691.588288  &  57691.600612  &  17.09  $\pm$  0.05   \\
            &       &  57691.655962  &  57691.657762  &  16.93  $\pm$  0.12   \\
            &       &  57691.723637  &  57691.731484  &  17.09  $\pm$  0.06   \\
            &       &  57691.791305  &  57691.793444  &  17.06  $\pm$  0.12   \\
            &       &  57691.862352  &  57691.863146  &  17.04  $\pm$  0.19   \\
 25-12-2019 & F148W &  58841.831743  &  58841.833213  &  18.61  $\pm$  0.13  \\
            & F154W &  58841.904312  &  58841.905736  &  18.55  $\pm$  0.15  \\
            & F169M &  58841.976881  &  58841.985370  &  18.11  $\pm$  0.06  \\
            &       &  58842.305330  &  58842.311886  &  18.17  $\pm$  0.07  \\
            & F172M &  58842.643654  &  58842.654596  &  17.91  $\pm$  0.09  \\ 
\enddata
\end{deluxetable*}

\begin{deluxetable}{ccccc}
\section{Brightness measurements of the source 1ES 0229+200}
\tablehead{\colhead{Date}  & \colhead{Filter} &  \colhead{UTstart} & \colhead{UTend} & \colhead{Mag}}
\startdata
 01-10-2017 & F154W &  58027.768812  &  58027.790965  &  20.43  $\pm$  0.10  \\
            &       &  58027.838336  &  58027.850038  &  20.35  $\pm$  0.10  \\
            &       &  58027.910905  &  58027.926301  &  20.44  $\pm$  0.12  \\
            & N245M &  58027.768763  &  58027.791014  &  19.77  $\pm$  0.05  \\
            &       &  58027.838287  &  58027.858677  &  19.84  $\pm$  0.06  \\
            &       &  58027.910856  &  58027.926351  &  19.89  $\pm$  0.07  \\ 
 09-12-2017 & F154W &  58096.737591  &  58096.743303  &  20.23  $\pm$  0.17  \\
            &       &  58096.789544  &  58096.810980  &  20.27  $\pm$  0.09  \\
            &       &  58096.859519  &  58096.878642  &  20.46  $\pm$  0.11  \\
            &       &  58096.932088  &  58096.944207  &  20.62  $\pm$  0.15  \\
            & N245M &  58096.737542  &  58096.743352  &  20.00  $\pm$  0.11  \\
            &       &  58096.789495  &  58096.811030  &  19.89  $\pm$  0.06  \\
            &       &  58096.859470  &  58096.878692  &  19.91  $\pm$  0.06  \\
            &       &  58096.932039  &  58096.944257  &  19.95  $\pm$  0.08  \\
 22-12-2017 & F154W &  58108.834869  &  58108.853240  &  20.44  $\pm$  0.11   \\
            &       &  58108.907438  &  58108.920891  &  20.35  $\pm$  0.12   \\
            &       &  58109.104261  &  58109.109884  &  20.42  $\pm$  0.19  \\
            &       &  58109.171930  &  58109.180995  &  20.42  $\pm$  0.15  \\
            & N245M &  58108.834820  &  58108.853290  &  19.89  $\pm$  0.06  \\
            &       &  58108.907389  &  58108.920941  &  19.99  $\pm$  0.07  \\
            &       &  58108.979959  &  58108.988579  &  19.89  $\pm$  0.09  \\
            &       &  58109.052528  &  58109.056241  &  19.99  $\pm$  0.14  \\
            &       &  58109.104212  &  58109.109934  &  19.91  $\pm$  0.11  \\
            &       &  58109.171881  &  58109.181045  &  20.03  $\pm$  0.09  \\          
 08-01-2018 & F154W &  58126.562901  &  58126.577813  &  20.36  $\pm$  0.12  \\
            &       &  58126.630575  &  58126.645463  &  20.34  $\pm$  0.11  \\
            &       &  58126.698249  &  58126.713114  &  20.40  $\pm$  0.12 \\
            &       &  58126.765924  &  58126.779650  &  20.13  $\pm$  0.11  \\
            & N245M &  58126.562852  &  58126.577862  &  19.86  $\pm$  0.07  \\
            &       &  58126.630526  &  58126.645513  &  19.78  $\pm$  0.06  \\
            &       &  58126.698201  &  58126.713163  &  19.87  $\pm$  0.07  \\
            &       &  58126.765875  &  58126.779700  &  19.70  $\pm$  0.06  \\ 
\enddata
\end{deluxetable}

\begin{deluxetable}{ccccc}
\section{Brightness measurements of the source OJ 287}
\tablehead{\colhead{Date}  & \colhead{Filter} &  \colhead{UTstart} & \colhead{UTend} & \colhead{Mag}}
\startdata
 10-04-2017 & F169M &  57852.921707  &  57852.931896  &  16.50  $\pm$  0.03  \\
            &       &  57852.996559  &  57853.001230  &  16.42  $\pm$  0.04  \\
            &       &  57853.124063  &  57853.127295  &  16.54  $\pm$  0.05  \\
            &       &  57853.193861  &  57853.200357  &  16.52  $\pm$  0.03  \\
            &       &  57853.263652  &  57853.273780  &  16.48  $\pm$  0.03  \\
            &       &  57853.333450  &  57853.347202  &  16.48  $\pm$  0.02  \\
            &       &  57853.407498  &  57853.420388  &  16.43  $\pm$  0.02  \\
            &       &  57853.473045  &  57853.483899  &  16.47  $\pm$  0.03  \\
            &       &  57853.542843  &  57853.551586  &  16.49  $\pm$  0.03  \\
            &       &  57853.612640  &  57853.629733  &  16.51  $\pm$  0.02  \\
            &       &  57853.682439  &  57853.686358  &  16.40  $\pm$  0.04  \\
            & N245M &  57852.921656  &  57852.931948  &  15.84  $\pm$  0.01  \\
            &       &  57852.997824  &  57853.001281  &  15.84  $\pm$  0.02  \\
            &       &  57853.124012  &  57853.127347  &  15.80  $\pm$  0.02  \\
            &       &  57853.193810  &  57853.200408  &  15.83  $\pm$  0.01  \\
            &       &  57853.263602  &  57853.273830  &  15.80  $\pm$  0.01  \\
            &       &  57853.333399  &  57853.347253  &  15.81  $\pm$  0.01  \\
            &       &  57853.409942  &  57853.415465  &  15.79  $\pm$  0.02  \\ 
            & N263M &  57853.417144  &  57853.420440  &  15.71  $\pm$  0.02  \\
            &       &  57853.472995  &  57853.483899  &  15.70  $\pm$  0.02  \\
            &       &  57853.542792  &  57853.551586  &  15.71  $\pm$  0.02  \\
            &       &  57853.612590  &  57853.629784  &  15.73  $\pm$  0.01  \\
            &       &  57853.682388  &  57853.686437  &  15.71  $\pm$  0.02  \\ 
 18-05-2020 & F148W  &  58984.390970  &  58984.396566  &  16.03  $\pm$  0.02  \\
            &        &  58984.458633  &  58984.469640  &  16.10  $\pm$  0.02  \\
            &        &  58984.526301  &  58984.537291  &  16.09  $\pm$  0.02  \\
            &        &  58984.593963  &  58984.604941  &  16.06  $\pm$  0.02  \\
            &        &  58984.661625  &  58984.664079  &  16.07  $\pm$  0.03  \\
            &        &  58984.729287  &  58984.732999  &  16.05  $\pm$  0.03  \\ 
            & F154W  &  58984.734732  &  58984.740242  &  15.95  $\pm$  0.02  \\
            &        &  58984.796957  &  58984.807898  &  16.01  $\pm$  0.02  \\
            &        &  58984.868199  &  58984.875548  &  15.96  $\pm$  0.02  \\  
            &        &  58984.940768  &  58984.943199  &  15.96  $\pm$  0.04  \\
            &        &  58985.202923  &  58985.213006  &  15.98  $\pm$  0.02  \\
            &        &  58985.270585  &  58985.281451  &  15.98  $\pm$  0.02  \\
            &        &  58985.338248  &  58985.349101  &  15.99  $\pm$  0.02  \\
            &        &  58985.405932  &  58985.407435  &  15.85  $\pm$  0.04  \\
            & F172M  &  58986.827577  &  58986.837416  &  15.70  $\pm$  0.04  \\
            &        &  58986.900145  &  58986.905066  &  15.68  $\pm$  0.05  \\
            &        &  58987.165157  &  58987.172383  &  15.62  $\pm$  0.04  \\
            &        &  58987.232825  &  58987.243306  &  15.58  $\pm$  0.03  \\
            &        &  58987.300488  &  58987.310956  &  15.54  $\pm$  0.03  \\
            &        &  58987.368156  &  58987.378613  &  15.61  $\pm$  0.04  \\
            &        &  58987.435818  &  58987.446263  &  15.60  $\pm$  0.04  \\  
            &        &  58987.503480  &  58987.505199  &  15.69  $\pm$  0.08  \\
            &        &  58987.638812  &  58987.649214  &  15.55  $\pm$  0.03  \\
            &        &  58987.706479  &  58987.710244  &  15.59  $\pm$  0.05  \\
            & F148Wa &  58985.409168  &  58985.416746  &  16.12  $\pm$  0.02  \\
            &        &  58985.473572  &  58985.481214  &  16.11  $\pm$  0.02  \\
            &        &  58985.608902  &  58985.619709  &  16.10  $\pm$  0.02  \\
            &        &  58985.676565  &  58985.687359  &  16.10  $\pm$  0.02  \\
            &        &  58985.744227  &  58985.748966  &  16.10  $\pm$  0.02  \\
            &        &  58986.556185  &  58986.561064  &  16.03  $\pm$  0.02  \\ 
\enddata
\end{deluxetable}

\begin{deluxetable}{ccccc}
\tablehead{\colhead{Date}  & \colhead{Filter} &  \colhead{UTstart} & \colhead{UTend} & \colhead{Mag}}
\startdata
18-04-2018 & F169M &  58225.694593  &  58225.707433  &  17.14  $\pm$  0.03   \\
           &       &  58225.758997  &  58225.775095  &  17.12  $\pm$  0.03   \\
           &       &  58225.827579  &  58225.842745  &  17.17  $\pm$  0.03   \\
           &       &  58225.900148  &  58225.910402  &  17.12  $\pm$  0.04   \\
           &       &  58225.972717  &  58225.977893  &  17.15  $\pm$  0.05   \\
           &       &  58226.232633  &  58226.240782  &  17.15  $\pm$  0.04   \\
           &       &  58226.300301  &  58226.316327  &  17.07  $\pm$  0.03   \\
           &       &  58226.367963  &  58226.383990  &  17.11  $\pm$  0.03   \\
           &       &  58226.435625  &  58226.451641  &  17.10  $\pm$  0.03   \\
           &       &  58226.503294  &  58226.519303  &  17.10  $\pm$  0.03   \\
           &       &  58226.570956  &  58226.586954  &  17.06  $\pm$  0.03   \\
           &       &  58226.638618  &  58226.654616  &  17.03  $\pm$  0.03   \\
           &       &  58226.706281  &  58226.722266  &  17.08  $\pm$  0.03   \\
           &       &  58226.773949  &  58226.789928  &  17.09  $\pm$  0.03   \\
           &       &  58226.843550  &  58226.857579  &  17.18  $\pm$  0.03   \\
           &       &  58226.916119  &  58226.925241  &  17.22  $\pm$  0.04   \\
           &       &  58226.988688  &  58226.992891  &  17.18  $\pm$  0.06   \\
           &       &  58227.179928  &  58227.183148  &  17.21  $\pm$  0.07   \\
           &       &  58227.247590  &  58227.259189  &  17.19  $\pm$  0.04   \\
           &       &  58227.315253  &  58227.331167  &  17.11  $\pm$  0.03   \\
           &       &  58227.382921  &  58227.398818  &  17.15  $\pm$  0.03   \\
           &       &  58227.450583  &  58227.466479  &  17.19  $\pm$  0.03   \\
           &       &  58227.518251  &  58227.534130  &  17.12  $\pm$  0.03   \\
           &       &  58227.585913  &  58227.596754  &  17.16  $\pm$  0.04   \\
           & F172M &  58223.458444  &  58223.461745  &  16.81  $\pm$  0.09   \\
           &       &  58223.526112  &  58223.542435  &  16.92  $\pm$  0.05   \\
           &       &  58223.593774  &  58223.610092  &  16.89  $\pm$  0.05   \\
           &       &  58223.661437  &  58223.677748  &  16.88  $\pm$  0.05   \\
           &       &  58223.729099  &  58223.745405  &  16.96  $\pm$  0.05   \\
           &       &  58223.796775  &  58223.813061  &  16.94  $\pm$  0.05   \\
           &       &  58223.867850  &  58223.880717  &  16.98  $\pm$  0.05   \\
           &       &  58223.940425  &  58223.948374  &  16.94  $\pm$  0.07   \\
           &       &  58224.202735  &  58224.206411  &  16.94  $\pm$  0.09   \\
           &       &  58224.270403  &  58224.281060  &  17.10  $\pm$  0.06   \\
           &       &  58224.338065  &  58224.354312  &  17.01  $\pm$  0.05   \\
           &       &  58224.405728  &  58224.421968  &  16.92  $\pm$  0.05   \\
           &       &  58224.473390  &  58224.489624  &  17.00  $\pm$  0.05   \\
           &       &  58224.541052  &  58224.557281  &  16.98  $\pm$  0.05   \\
           &       &  58224.608720  &  58224.624937  &  17.04  $\pm$  0.05   \\
           &       &  58224.676383  &  58224.692593  &  16.94  $\pm$  0.05   \\
           &       &  58224.811721  &  58224.827906  &  16.97  $\pm$  0.05   \\
           &       &  58224.884171  &  58224.895546  &  16.99  $\pm$  0.06   \\
           &       &  58224.956740  &  58224.963219  &  17.06  $\pm$  0.08   \\
           &       &  58225.217688  &  58225.223425  &  16.92  $\pm$  0.07   \\
           &       &  58225.285349  &  58225.299468  &  16.94  $\pm$  0.05   \\
           &       &  58225.353011  &  58225.369157  &  16.95  $\pm$  0.05   \\
           &       &  58225.420680  &  58225.436807  &  17.03  $\pm$  0.05   \\
           &       &  58225.488342  &  58225.504464  &  16.99  $\pm$  0.05   \\
           &       &  58225.556004  &  58225.572126  &  16.98  $\pm$  0.05   \\
           &       &  58225.623666  &  58225.632778  &  16.90  $\pm$  0.06   \\
           &       &  58225.691329  &  58225.692828  &  17.02  $\pm$  0.15   \\
\enddata
\end{deluxetable}

\begin{deluxetable}{cccccc}
\section{Brightness measurements of the source 1ES 1101$-$232}
\tablehead{\colhead{Date}  & \colhead{Filter} &  \colhead{UTstart} & \colhead{UTend} & \colhead{Mag}}
\startdata
  30-12-2016 & F154W &  57750.400520  & 57750.416409  &  18.39   $\pm$  0.04  \\
             &       &  57750.468182  & 57750.484078  &  18.44   $\pm$  0.04  \\
             &       &  57750.535845  & 57750.551746  &  18.39   $\pm$  0.04  \\
             &       &  57750.603507  & 57750.619414  &  18.42   $\pm$  0.04  \\
             &       &  57750.671163  & 57750.687082  &  18.42   $\pm$  0.04  \\
             &       &  57750.738825  & 57750.754751  &  18.46   $\pm$  0.04  \\
             &       &  57750.806482  & 57750.822419  &  18.52   $\pm$  0.05  \\
             &       &  57750.874144  & 57750.890087  &  18.46   $\pm$  0.04  \\
             &       &  57750.944243  & 57750.957755  &  18.44   $\pm$  0.05  \\
             &       &  57751.016812  & 57751.025423  &  18.51   $\pm$  0.06  \\
             &       &  57751.089381  & 57751.093092  &  18.41   $\pm$  0.09  \\
             &       &  57751.144787  & 57751.147032  &  18.71   $\pm$  0.13  \\
             &       &  57751.212450  & 57751.219254  &  18.51   $\pm$  0.07  \\
             &       &  57751.280106  & 57751.291130  &  18.43   $\pm$  0.05  \\
             &       &  57751.347768  & 57751.362313  &  18.44   $\pm$  0.05  \\
             &       &  57751.415431  & 57751.428900  &  18.43   $\pm$  0.04  \\
             &       &  57751.483093  & 57751.499101  &  18.49   $\pm$  0.05  \\
             &       &  57751.550749  & 57751.566769  &  18.44   $\pm$  0.04  \\
             &       &  57751.618411  & 57751.634437  &  18.39   $\pm$  0.04  \\
             &       &  57751.686074  & 57751.702111  &  18.40   $\pm$  0.04  \\
             &       &  57751.753730  & 57751.769779  &  18.34   $\pm$  0.04  \\
             &       &  57751.821392  & 57751.837447  &  18.47   $\pm$  0.04  \\
             &       &  57751.889056  & 57751.905110  &  18.40   $\pm$  0.04  \\
             &       &  57751.960558  & 57751.972778  &  18.43   $\pm$  0.05  \\
             &       &  57752.033133  & 57752.040446  &  18.44   $\pm$  0.06  \\
             &       &  57752.105702  & 57752.108114  &  18.38	 $\pm$  0.11  \\
             &       &  57752.159686  & 57752.163003  &  18.29   $\pm$  0.09  \\
             &       &  57752.229222  & 57752.235231  &  18.40   $\pm$  0.07  \\
             &       &  57752.295010  & 57752.306408  &  18.45   $\pm$  0.05  \\
             &       &  57752.362673  & 57752.377590  &  18.43   $\pm$  0.05  \\
             &       &  57752.430329  & 57752.446455  &  18.40   $\pm$  0.04  \\
             &       &  57752.497991  & 57752.501664  &  18.55   $\pm$  0.10  \\
             & N263M &  57750.400471  & 57750.416460  &  17.93   $\pm$  0.03  \\
             &       &  57750.468133  & 57750.484127  &  17.89   $\pm$  0.03  \\
             &       &  57750.535796  & 57750.551795  &  17.88   $\pm$  0.03  \\
             &       &  57750.603458  & 57750.619464  &  17.84   $\pm$  0.03  \\
             &       &  57750.671114  & 57750.687132  &  17.91   $\pm$  0.03  \\
             &       &  57750.738777  & 57750.754800  &  17.92   $\pm$  0.03  \\
             &       &  57750.806433  & 57750.822468  &  17.90   $\pm$  0.03  \\
             &       &  57750.874095  & 57750.890136  &  17.90   $\pm$  0.03  \\
             &       &  57750.944195  & 57750.957805  &  17.91   $\pm$  0.03  \\
             &       &  57751.016764  & 57751.025473  &  17.91   $\pm$  0.04  \\
             &       &  57751.089333  & 57751.093141  &  17.89   $\pm$  0.06  \\
             &       &  57751.144739  & 57751.147082  &  17.92   $\pm$  0.07  \\
             &       &  57751.212401  & 57751.219304  &  17.92   $\pm$  0.04  \\
             &       &  57751.280057  & 57751.291180  &  17.85   $\pm$  0.03  \\
             &       &  57751.347719  & 57751.362362  &  17.90   $\pm$  0.03  \\
             &       &  57751.415382  & 57751.431482  &  17.92   $\pm$  0.03  \\
             &       &  57751.483044  & 57751.499150  &  17.94   $\pm$  0.03  \\
             &       &  57751.550700  & 57751.566818  &  17.93   $\pm$  0.03  \\
             &       &  57751.618363  & 57751.634487  &  17.91   $\pm$  0.03  \\
             &       &  57751.686025  & 57751.702161  &  17.93   $\pm$  0.03  \\
             &       &  57751.753681  & 57751.769829  &  17.96   $\pm$  0.03  \\
             &       &  57751.821343  & 57751.837497  &  17.93   $\pm$  0.03  \\
\enddata
\end{deluxetable}

\begin{deluxetable}{ccccc}
\tablehead{\colhead{Date}  & \colhead{Filter} &  \colhead{UTstart} & \colhead{UTend} & \colhead{Mag}}
\startdata             
             &       &  57751.889007  & 57751.905159  &  17.93   $\pm$  0.03  \\
             &       &  57751.960509  & 57751.972827  &  17.95   $\pm$  0.03  \\
             &       &  57752.033084  & 57752.040496  &  17.98   $\pm$  0.04  \\
             &       &  57752.105653  & 57752.108164  &  17.82   $\pm$  0.07  \\
             &       &  57752.159637  & 57752.163053  &  18.05	 $\pm$  0.07  \\
             &       &  57752.227299  & 57752.235281  &  17.95   $\pm$  0.04  \\
             &       &  57752.294961  & 57752.306457  &  17.86   $\pm$  0.03  \\
             &       &  57752.362624  & 57752.377640  &  17.96   $\pm$  0.03  \\
             &       &  57752.430280  & 57752.446505  &  17.90   $\pm$  0.03  \\
             &       &  57752.497942  & 57752.501714  &  18.04   $\pm$  0.06  \\
\enddata
\end{deluxetable}

\begin{deluxetable}{ccccc}
\section{Brightness measurements of the source 1ES 1218+304}
\tablehead{\colhead{Date}  & \colhead{Filter} &  \colhead{UTstart} & \colhead{UTend} & \colhead{Mag}}
\startdata
  21-05-2016  & F148W &  57528.302435  &  57528.308327  &  17.88  $\pm$  0.05 \\
              &       &  57528.372248  &  57528.383183  &  17.91  $\pm$  0.04 \\
              &       &  57528.442067  &  57528.458041  &  17.99  $\pm$  0.03 \\
              &       &  57528.511879  &  57528.529370  &  17.93  $\pm$  0.03 \\
              &       &  57528.581692  &  57528.599164  &  17.88  $\pm$  0.03 \\
              &       &  57528.651504  &  57528.668952  &  17.90  $\pm$  0.03 \\
              &       &  57528.721317  &  57528.738740  &  17.92  $\pm$  0.03 \\
              &       &  57528.791136  &  57528.808534  &  17.90  $\pm$  0.03 \\
              &       &  57528.861914  &  57528.878322  &  17.96  $\pm$  0.03 \\
              &       &  57528.936769  &  57528.948117  &  17.94  $\pm$  0.04 \\
              &       &  57529.011625  &  57529.017904  &  17.91  $\pm$  0.05 \\
              &       &  57529.279823  &  57529.281816  &  17.82  $\pm$  0.08 \\
              &       &  57529.419509  &  57529.431526  &  17.90  $\pm$  0.03 \\
              & N245M &  57528.302745  &  57528.308378  &  17.41  $\pm$  0.03 \\
              &       &  57528.373033  &  57528.383234  &  17.36  $\pm$  0.02 \\
              &       &  57528.442016  &  57528.458092  &  17.38  $\pm$  0.02 \\
              &       &  57528.511829  &  57528.529421  &  17.36  $\pm$  0.02 \\
              &       &  57528.581641  &  57528.599215  &  17.39  $\pm$  0.02 \\
              &       &  57528.651454  &  57528.669003  &  17.36  $\pm$  0.02 \\
              &       &  57528.721266  &  57528.738792  &  17.38  $\pm$  0.02 \\
              &       &  57528.791085  &  57528.808586  &  17.38  $\pm$  0.02 \\
              &       &  57528.861863  &  57528.878374  &  17.38  $\pm$  0.02 \\
              &       &  57528.936719  &  57528.948168  &  17.39  $\pm$  0.02 \\
              &       &  57529.011574  &  57529.017956  &  17.36  $\pm$  0.03 \\
              &       &  57529.279773  &  57529.281867  &  17.42  $\pm$  0.05 \\
              &       &  57529.419913  &  57529.431577  &  17.35  $\pm$  0.02 \\ 
\enddata
\end{deluxetable}

\begin{deluxetable}{ccccc}
\section{Brightness measurements of the source H 1426+428}
\tablehead{\colhead{Date}  & \colhead{Filter} &  \colhead{UTstart} & \colhead{UTend} & \colhead{Mag}}
\startdata
  05-03-2018  & F148W &  58181.865544  &  58181.867427  & 18.47  $\pm$ 0.11 \\
              &       &  58181.919243  &  58181.935090  & 18.65  $\pm$ 0.04 \\
              &       &  58181.991811  &  58181.998758  & 18.60  $\pm$ 0.06 \\
              &       &  58182.064381  &  58182.070426  & 18.53  $\pm$ 0.06 \\
              &       &  58182.253915  &  58182.261613  & 18.63  $\pm$ 0.06 \\
              &       &  58182.321565  &  58182.337656  & 18.64  $\pm$ 0.04 \\
              &       &  58182.389210  &  58182.408744  & 18.63  $\pm$ 0.04 \\
              &       &  58182.456859  &  58182.463086  & 18.66  $\pm$ 0.07 \\
              & F154W &  58181.577434  &  58181.596767  & 18.58  $\pm$ 0.04 \\
              &       &  58181.645079  &  58181.664441  & 18.50  $\pm$ 0.04 \\
              &       &  58181.712729  &  58181.732103  & 18.56  $\pm$ 0.04 \\
              &       &  58181.780374  &  58181.799771  & 18.47  $\pm$ 0.04 \\
              &       &  58181.848025  &  58181.863807  & 18.52  $\pm$ 0.05 \\
              & N242W &  58181.577385  &  58181.596816  & 18.13  $\pm$ 0.01 \\
              &       &  58181.645030  &  58181.664490  & 18.14  $\pm$ 0.01 \\
              &       &  58181.712680  &  58181.732152  & 18.18  $\pm$ 0.01 \\
              &       &  58181.780325  &  58181.781955  & 18.21  $\pm$ 0.05 \\          
              & N245M &  58181.783577  &  58181.799821  & 18.18  $\pm$ 0.03 \\
              &       &  58181.847977  &  58181.867477  & 18.19  $\pm$ 0.03 \\
              &       &  58181.919194  &  58181.935139  & 18.23  $\pm$ 0.03 \\
              &       &  58181.991762  &  58181.998808  & 18.16  $\pm$ 0.04 \\
              &       &  58182.064332  &  58182.070476  & 18.19  $\pm$ 0.04 \\
              &       &  58182.253866  &  58182.261659  & 18.16  $\pm$ 0.04 \\
              &       &  58182.321516  &  58182.337706  & 18.20  $\pm$ 0.03 \\
              &       &  58182.389161  &  58182.408793  & 18.15  $\pm$ 0.02 \\
              &       &  58182.456810  &  58182.463086  & 18.25	 $\pm$ 0.05 \\
\enddata
\end{deluxetable}

\begin{deluxetable}{cccccc}
\section{Brightness measurements of the source PKS 1510$-$089}
\tablehead{\colhead{Date}  & \colhead{Filter} &  \colhead{UTstart} & \colhead{UTend} & \colhead{Mag}}
\startdata
 30-03-2016  & F172M &  57476.398455  &  57476.400733  & 17.30   $\pm$  0.14  \\
             &       &  57476.455977  &  57476.473306  & 17.12   $\pm$  0.05  \\
             &       &  57476.523622  &  57476.542681  & 17.09   $\pm$  0.05  \\
             &       &  57476.591272  &  57476.610338  & 17.06   $\pm$  0.05  \\
             &       &  57476.658923  &  57476.678000  & 17.09   $\pm$  0.05  \\
             &       &  57476.726567  &  57476.745657  & 17.06   $\pm$  0.05  \\
             &       &  57476.794218  &  57476.813313  & 17.05   $\pm$  0.05  \\
             &       &  57476.864493  &  57476.880975  & 17.13   $\pm$  0.05  \\
             &       &  57476.937062  &  57476.948632  & 17.08   $\pm$  0.06  \\
             &       &  57477.082206  &  57477.083941  & 17.02   $\pm$  0.14  \\
             &       &  57477.335410  &  57477.344146  & 17.16   $\pm$  0.07  \\
             &       &  57477.403060  &  57477.416716  & 17.16   $\pm$  0.06  \\
             &       &  57477.470711  &  57477.489285  & 17.08   $\pm$  0.05  \\
             &       &  57477.538361  &  57477.557557  & 17.06   $\pm$  0.05  \\
             &       &  57477.606018  &  57477.625214  & 17.09   $\pm$  0.05  \\
             &       &  57477.673656  &  57477.692876  & 17.11   $\pm$  0.05  \\
             &       &  57477.741307  &  57477.760532  & 17.09   $\pm$  0.05  \\
             &       &  57477.808957  &  57477.828195  & 17.10   $\pm$  0.05  \\
             &       &  57477.880471  &  57477.895851  & 17.09   $\pm$  0.05  \\      
             &       &  57477.953040  &  57477.963508  & 17.15   $\pm$  0.06  \\       
             &       &  57478.025609  &  57478.031161  & 17.05   $\pm$  0.08  \\       
             & N219M &  57476.398407  &  57476.400783  & 17.33   $\pm$  0.11  \\
             &       &  57476.455928  &  57476.473356  & 17.27   $\pm$  0.04  \\
             &       &  57476.523573  &  57476.542731  & 17.34   $\pm$  0.04  \\
             &       &  57476.591223  &  57476.610388  & 17.26   $\pm$  0.04  \\
             &       &  57476.658874  &  57476.678049  & 17.38   $\pm$  0.05  \\
             &       &  57476.726518  &  57476.745706  & 17.28   $\pm$  0.04  \\
             &       &  57476.794169  &  57476.813362  & 17.28   $\pm$  0.04  \\
             &       &  57476.864444  &  57476.881025  & 17.30   $\pm$  0.05  \\
             &       &  57476.937013  &  57476.948681  & 17.31   $\pm$  0.05  \\
             &       &  57477.082157  &  57477.083991  & 17.42   $\pm$  0.13  \\
             &       &  57477.335361  &  57477.344196  & 17.27   $\pm$  0.06  \\
             &       &  57477.403011  &  57477.416765  & 17.37   $\pm$  0.05  \\
             &       &  57477.470662  &  57477.489334  & 17.31   $\pm$  0.04  \\       
             &       &  57477.538312  &  57477.557607  & 17.36   $\pm$  0.04  \\
             &       &  57477.605967  &  57477.625263  & 17.37   $\pm$  0.04  \\
             &       &  57477.673607  &  57477.692926  & 17.38   $\pm$  0.04  \\
             &       &  57477.741258  &  57477.760582  & 17.32   $\pm$  0.04  \\
             &       &  57477.808908  &  57477.828244  & 17.34   $\pm$  0.04  \\
             &       &  57477.880422  &  57477.895901  & 17.43   $\pm$  0.05  \\
             &       &  57477.952992  &  57477.963557  & 17.32   $\pm$  0.06  \\
             &       &  57478.025561  &  57478.031211  & 17.42   $\pm$  0.08  \\
 16-03-2018  & F172M &  58192.878269  &  58192.896628  & 17.30   $\pm$  0.05  \\
             &       &  58192.950836  &  58192.964290  & 17.30   $\pm$  0.06  \\
             &       &  58193.023413  &  58193.031953  & 17.24   $\pm$  0.07  \\
             &       &  58193.095982  &  58193.099615  & 17.14   $\pm$  0.10  \\
             &       &  58193.215767  &  58193.225855  & 17.30   $\pm$  0.07  \\
             &       &  58193.351074  &  58193.369257  & 17.17   $\pm$  0.05  \\
             &       &  58193.418729  &  58193.437939  & 17.26   $\pm$  0.05  \\
             &       &  58193.486381  &  58193.505601  & 17.30   $\pm$  0.05  \\
             &       &  58193.554037  &  58193.573275  & 17.26   $\pm$  0.05  \\
             &       &  58193.621688  &  58193.640937  & 17.35   $\pm$  0.05  \\
             &       &  58193.689344  &  58193.695960  & 17.36   $\pm$  0.08  \\
\enddata
\end{deluxetable}

\begin{deluxetable}{ccccc}
\tablehead{\colhead{Date}  & \colhead{Filter} & \colhead{ UTstart} & \colhead{UTend} & \colhead{Mag}}
\startdata
             & N219M &  58192.878220  &  58192.896677  & 17.55   $\pm$  0.05  \\
             &       &  58192.950788  &  58192.964340  & 17.55   $\pm$  0.06  \\
             &       &  58193.023364  &  58193.032002  & 17.43   $\pm$  0.06  \\
             &       &  58193.095933  &  58193.099664  & 17.93   $\pm$  0.12  \\
             &       &  58193.215718  &  58193.225905  & 17.53   $\pm$  0.06  \\
             &       &  58193.351025  &  58193.369307  & 17.53   $\pm$  0.05  \\
             &       &  58193.418680  &  58193.437988  & 17.48   $\pm$  0.05  \\
             &       &  58193.486332  &  58193.505650  & 17.58   $\pm$  0.05  \\
             &       &  58193.553989  &  58193.573325  & 17.64   $\pm$  0.05  \\
             &       &  58193.621639  &  58193.640987  & 17.55   $\pm$  0.05  \\
             &       &  58193.689295  &  58193.696009  & 17.63   $\pm$  0.08  \\   
 15-06-2018  & F172M &  58284.086589  &  58284.089744  & 17.15   $\pm$  0.11  \\
             &       &  58284.154258  &  58284.159266  & 17.42   $\pm$  0.10  \\
             &       &  58284.297771  &  58284.306408  & 17.33   $\pm$  0.07  \\
             &       &  58284.357274  &  58284.377591  & 17.29   $\pm$  0.05  \\
             &       &  58284.424942  &  58284.445639  & 17.32   $\pm$  0.05  \\
             &       &  58284.492611  &  58284.513314  & 17.34   $\pm$  0.05  \\
             &       &  58284.560285  &  58284.580982  & 17.28   $\pm$  0.05  \\
             &       &  58284.627955  &  58284.648656  & 17.28   $\pm$  0.05  \\
             &       &  58284.695628  &  58284.716318  & 17.29   $\pm$  0.05  \\
             &       &  58284.763296  &  58284.771371  & 17.26   $\pm$  0.07  \\ 
\enddata
\end{deluxetable}

\begin{deluxetable}{ccccc}
\section{Brightness measurements of the source M\lowercase{rk} 501}
\tablehead{\colhead{Date}  & \colhead{Filter} &  \colhead{UTstart} & \colhead{UTend} & \colhead{Mag}}
\startdata
 15-08-2016 & F154W &  57615.260263  &  57615.269143  & 15.82   $\pm$  0.02  \\
            &       &  57615.323334  &  57615.337866  & 15.83   $\pm$  0.02  \\
            &       &  57615.391014  &  57615.405528  & 15.84   $\pm$  0.02  \\
            &       &  57615.458688  &  57615.473184  & 15.85   $\pm$  0.02  \\
            &       &  57615.526368  &  57615.540841  & 15.84   $\pm$  0.02  \\
            &       &  57615.594042  &  57615.608497  & 15.84   $\pm$  0.02  \\
            &       &  57615.661723  &  57615.676160  & 15.86   $\pm$  0.02  \\
            &       &  57615.729396  &  57615.743816  & 15.85   $\pm$  0.02  \\
            &       &  57615.797070  &  57615.811472  & 15.84   $\pm$  0.02  \\
            &       &  57615.867618  &  57615.879134  & 15.87   $\pm$  0.02  \\
            &       &  57615.940187  &  57615.946790  & 15.84   $\pm$  0.02  \\
            & N219M &  57615.255611  &  57615.269193  & 15.54   $\pm$  0.03  \\
            &       &  57615.323285  &  57615.337916  & 15.57   $\pm$  0.03  \\
            &       &  57615.390965  &  57615.405578  & 15.54   $\pm$  0.03  \\
            &       &  57615.458639  &  57615.473234  & 15.54   $\pm$  0.03  \\
            &       &  57615.526319  &  57615.540891  & 15.58   $\pm$  0.03  \\
            &       &  57615.593993  &  57615.608547  & 15.56   $\pm$  0.03  \\
            &       &  57615.661675  &  57615.676209  & 15.53   $\pm$  0.03  \\
            &       &  57615.729347  &  57615.743866  & 15.53   $\pm$  0.03  \\
            &       &  57615.797021  &  57615.811522  & 15.59   $\pm$  0.03  \\
            &       &  57615.867569  &  57615.879184  & 15.57   $\pm$  0.03  \\
            &       &  57615.940138  &  57615.946840  & 15.50   $\pm$  0.03  \\
 28-03-2020 & F148W &  58935.275928  &  58935.281755  & 16.09   $\pm$  0.02  \\
            &       &  58935.343573  &  58935.355761  & 16.05   $\pm$  0.02  \\
            &       &  58935.817072  &  58935.833668  & 16.08   $\pm$  0.02  \\
            &       &  58935.884736  &  58935.889611  & 16.08   $\pm$  0.02  \\
            &       &  58936.028618  &  58936.036643  & 16.11   $\pm$  0.02  \\
            &       &  58936.101187  &  58936.104294  & 16.07   $\pm$  0.03  \\
            &       &  58936.290578  &  58936.297587  & 16.06   $\pm$  0.02  \\
            &       &  58936.493506  &  58936.510244  & 16.11	$\pm$  0.02  \\
            &       &  58936.561150  &  58936.577906  & 16.09   $\pm$  0.02  \\
            &       &  58936.628789  &  58936.645556  & 16.10   $\pm$  0.02  \\
            &       &  58936.696433  &  58936.713219  & 16.09   $\pm$  0.02  \\
            &       &  58936.764078  &  58936.780881  & 16.10   $\pm$  0.02  \\
            &       &  58936.831716  &  58936.848530  & 16.08   $\pm$  0.02  \\
            &       &  58936.899451  &  58936.916194  & 16.08   $\pm$  0.02  \\
            &       &  58936.972020  &  58936.983856  & 16.08   $\pm$  0.02  \\
            &       &  58937.044589  &  58937.051507  & 16.11   $\pm$  0.02  \\
            &       &  58937.117158  &  58937.119169  & 16.10   $\pm$  0.04  \\
            &       &  58937.237571  &  58937.241479  & 16.05   $\pm$  0.03  \\
            &       &  58937.305216  &  58937.317871  & 16.11   $\pm$  0.02  \\
            &       &  58937.372861  &  58937.389794  & 16.11   $\pm$  0.02  \\
            &       &  58937.440499  &  58937.457457  & 16.08   $\pm$  0.02  \\
            &       &  58937.508144  &  58937.525106  & 16.09   $\pm$  0.02  \\
            &       &  58937.575788  &  58937.592769  & 16.10   $\pm$  0.02  \\
            &       &  58937.643427  &  58937.660426  & 16.09   $\pm$  0.02  \\
            &       &  58937.711065  &  58937.726873  & 16.10   $\pm$  0.02 \\ 
\enddata
\end{deluxetable}

\begin{deluxetable}{ccccc}
\section{Brightness measurements of the source PKS 2155$-$304}
\tablehead{\colhead{Date}  & \colhead{Filter} &  \colhead{UTstart} & \colhead{UTend} & \colhead{Mag}}
\startdata
   21-05-2018   & F154W & 58259.329918  & 58259.332444   & 15.25  $\pm$  0.03     \\
                &       & 58259.397568  & 58259.405364   & 15.31  $\pm$  0.02     \\
                &       & 58259.465217  & 58259.478016   & 15.29  $\pm$  0.01     \\
                &       & 58259.532863  & 58259.544023   & 15.29  $\pm$  0.02     \\
   24-06-2018   & F154W & 58293.495067  & 58293.513388   & 15.17  $\pm$  0.01     \\
                &       & 58293.562712  & 58293.579373   & 15.17  $\pm$  0.01     \\ 
   16-09-2019   & F154W & 58742.573283  & 58742.593529	 & 15.02  $\pm$  0.01     \\
                &       & 58742.639023  & 58742.643320   & 15.04  $\pm$  0.02     \\
                &       & 58742.774366  & 58742.777812   & 14.99  $\pm$  0.02     \\
\enddata
\end{deluxetable}

\begin{deluxetable}{ccccc}
\section{Brightness measurements of the source 1ES 2344+514}
\tablehead{\colhead{Date}  & \colhead{Filter} &  \colhead{UTstart} & \colhead{UTend} & \colhead{Mag}}
\startdata
 06-06-2017 & F172M &  57910.717527  &  57910.727655  & 18.71  $\pm$  0.13  \\
            &       &  57910.787165  &  57910.797305  & 18.44  $\pm$  0.11  \\
            &       &  57910.856803  &  57910.863068  & 20.05  $\pm$  0.34  \\
            & N245M &  57910.717477  &  57910.727707  & 18.23  $\pm$  0.04  \\
            &       &  57910.787115  &  57910.797356  & 18.18  $\pm$  0.04  \\
            &       &  57910.856752  &  57910.860639  & 18.19  $\pm$  0.06  \\
 09-07-2017 & F172M &  57943.391924  &  57943.405146  & 18.36  $\pm$  0.10  \\
            &       &  57943.464512  &  57943.479844  & 18.54  $\pm$  0.10  \\
            &       &  57943.531900  &  57943.546931  & 18.58  $\pm$  0.10  \\
            & N245M &  57943.391873  &  57943.405197  & 18.25  $\pm$  0.03  \\
            &       &  57943.465318  &  57943.479895  & 18.18  $\pm$  0.03  \\
            &       &  57943.531849  &  57943.549083  & 18.18  $\pm$  0.03  \\
 07-08-2017 & F172M &  57972.282651  &  57972.300458  & 18.60  $\pm$  0.10  \\
            &       &  57972.352620  &  57972.374054  & 18.53  $\pm$  0.08  \\
            &       &  57972.422583  &  57972.430529  & 18.45  $\pm$  0.13  \\
            & N245M &  57972.282601  &  57972.300509  & 18.11  $\pm$  0.03  \\
            &       &  57972.352569  &  57972.374105  & 18.10  $\pm$  0.02  \\
            &       &  57972.422532  &  57972.428121  & 18.12  $\pm$  0.05  \\
 07-09-2017 & F172M &  58003.815110  &  58003.837802  & 18.87  $\pm$  0.10  \\
            &       &  58003.889949  &  58003.907581  & 19.02  $\pm$  0.12  \\
            &       &  58003.964788  &  58003.972611  & 19.13  $\pm$  0.19  \\
            & N245M &  58003.815060  &  58003.837853  & 18.60  $\pm$  0.03  \\
            &       &  58003.889899  &  58003.907632  & 18.65  $\pm$  0.03  \\
            &       &  58003.964738  &  58003.972663  & 18.63  $\pm$  0.05  \\
 22-11-2017 & N245M &  58078.933083  &  58078.942454  & 18.56  $\pm$  0.04 \\
            &       &  58079.005652  &  58079.010116  & 18.59  $\pm$  0.06 \\
            &       &  58079.127772  &  58079.130418  & 18.59  $\pm$  0.08 \\
            &       &  58079.195446  &  58079.201251  & 18.48  $\pm$  0.05 \\ 
 07-12-2017 & F172M &  58094.972718  &  58094.978781  & 18.62  $\pm$  0.16 \\
            &       &  58095.233275  &  58095.243561  & 19.03  $\pm$  0.15 \\
            & N245M &  58094.972670  &  58094.978831  & 18.40  $\pm$  0.05 \\
            &       &  58095.233226  &  58095.243611  & 18.37  $\pm$  0.04 \\           
 12-09-2018 & F172M &  58372.864037  &  58372.882670  & 18.81  $\pm$  0.10 \\
            &       &  58372.936606  &  58372.950332  & 18.58  $\pm$  0.11 \\
            &       &  58373.009175  &  58373.017994  & 18.78  $\pm$  0.15 \\
            &       &  58373.063674  &  58373.064969  & 18.73  $\pm$  0.37 \\            
\enddata
\end{deluxetable}

\end{document}